\documentclass[preprint2]{emulateapj}

\def\gsim{\;\lower4pt\hbox{${\buildrel\displaystyle >\over\sim}$}\;}
\def\lsim{\;\lower4pt\hbox{${\buildrel\displaystyle <\over\sim}$}\;}
\def\grls{\;\lower4pt\hbox{${\buildrel\displaystyle >\over <}$}\;}

\slugcomment{Received 2010...; accepted 2011...; published...}

\shorttitle{EECC for Star-Forming Cloud Core L1517B}
\shortauthors{Fu, Gao and Lou}

\begin{document}


\title{Starless Cloud Core L1517B in
Envelope Expansion with Core Collapse}


\author{Tian-Ming Fu\altaffilmark{1},{Yang Gao\altaffilmark{1,2}},
Yu-Qing Lou\altaffilmark{1,3,4}}


\altaffiltext{1}{Department of Physics and Tsinghua Center for
Astrophysics (THCA), Tsinghua University, Beijing 100084, China.}
\altaffiltext{2}{Center for Combustion Energy and Department of
Thermal Engineering, Tsinghua University, Beijing 100084, China.}
\altaffiltext{3}{Department of Astronomy and Astrophysics, the
University of Chicago, 5460 South Ellis Avenue, Chicago, IL 60637,
USA.} \altaffiltext{4}{National Astronomical Observatories,
 Chinese Academy of Science, A20, Datun Road, Beijing 100012, China.}


\begin{abstract}
Various spectral emission lines from star-forming molecular cloud
  core L1517B manifest red asymmetric double-peaked profiles with
  stronger red peaks and weaker blue peaks, in contrast to the
  oft-observed blue-skewed molecular spectral line profiles
  with blue peaks stronger than red peaks.
Invoking a spherically symmetric general polytropic hydrodynamic
  shock model for the envelope expansion with core collapse (EECC)
  phase, we show the radial flow velocity, mass density and
  temperature structures of self-similar evolution for L1517B
  in a dynamically consistent manner.
By prescribing simple radial profiles of abundance distribution
  for pertinent molecules, we perform molecular excitation and
  radiative transfer calculations using the publicly available
  RATRAN code set for the spherically symmetric case.
Emphatically, spectral profiles of line emissions from the same
  molecules but for different line transitions as well as spectra
  of closely pertinent isotopologues strongly constrain the
  self-similar hydrodynamics of a cloud core with prescribed
  abundances.
Our computational results show that the EECC model reproduces
  molecular spectral line profiles in sensible agreement with
  observational data of Institut de Radioastronomie
  Millim\'etrique (IRAM), Five College Radio Astronomical
  Observatory (FCRAO) and Effelsberg 100 m telescopes for L1517B.
We also report spatially resolved observations of
  optically thick line HCO$^+(1-0)$ using the Purple Mountain
  Observatory (PMO) 13.7 m telescope at Delingha in
  China and the relevant fitting results.
Hyperfine line structures of NH$_3$ and N$_2$H$^+$
  transitions are also fitted to consistently reveal
  the dynamics of central core collapse.
As a consistent model check, radial profiles of 1.2 mm and 850
  $\mu$m dust continua observed by IRAM 30 m telescope and the
  Submillimeter Common-User Bolometer Array (SCUBA), respectively,
  are also fitted numerically using the same EECC model that
  produces the molecular line profiles.
L1517B is likely undergoing an EECC shock phase.
For future observational tests,
  we also predict several molecular line profiles with spatial
  distributions, radial profile of sub-millimeter continuum at
  wavelength 450$\mu$m, as well as the radial profiles of the
  column density and visual extinction for L1517B.
\end{abstract}





\section{Introduction}{\label{sec:level1}}

Observations over past two decades provide effective diagnostics
  to examine dynamic properties of molecular clouds where
  star formations are taking place in relatively early phases.
These observations, including molecular spectral line profiles
  \citep[e.g.][]{bensonmyers1989,tafalla2002,aguti2007},
  radial profiles of (sub)millimeter radio continua
  \citep[e.g.][]{motte1998,shirley2000,stanke2006,nutterthompson2007},
  and infrared dust extinctions \citep[e.g.][]{alves2001}, can be
  utilized separately and/or in combination to reveal several key
  hydrodynamic cloud features, such as core collapses, envelope
  expansions, travelling shocks and turbulence in the early
  evolution of star-forming cloud cores.

On a relatively independent track, the theoretical model framework
  based on self-similar hydrodynamic collapses has achieved
  substantial development since late 1960s
  \citep[e.g.][]{bodenheimer,larson1969,penston1969,shu1977,hunter1977},
  and different variations of these theoretical model
  applications have been advanced in recent years
  \citep[e.g.][]{terebey1984,loushen2004,wanglou2008,stahleryen2009}.
For grossly spherical molecular clouds, the gas flow
  velocity, mass density and thermal temperature structures of
  star-forming molecular cloud cores predicted by theoretical
  models can be tested by several complementary
  high-resolution observations of selected source candidates.
Attempts of this kind have been pursued all along by comparing
  theoretical cloud model results with the observed molecular
  spectral line profiles
  \citep[e.g.][]{zhou1993,testisargent1998,tafalla2004,tafalla2006},
  (sub)millimeter radio continuum emissions
  \citep[e.g.][]{shirley2002,motteandre2001,harvey2003}, and
  infrared dust extinctions \citep[e.g.][]{alves2001,kandori2005}.

Molecular spectral line profiles are important to study physical
  structures and kinematics for various phases in star
  formation processes \citep[e.g.][]{dysonwilliams1997}.
They need to be physically interpreted and can be properly
  utilized to probe gas densities and temperatures through
  excitations of molecules, and detect cloud turbulence as
  well as systematic bulk flow motions through their line
  widths and Doppler shifts etc.
Observations of spectral lines more frequently exhibit blue skewed
  double-peak molecular profiles with stronger blue peaks and weaker
  red peaks \citep[i.e. the so-called blue profiles, see e.g.]
  []{gregersen1997,gregersen2000,lee1999}.
These blue asymmetric spectral line profiles of molecular
  transitions are usually interpreted as signatures of core
  collapses with a static outer envelope and a temperature variation
  involved \citep[e.g.][]{zhou1993,myers2000,gaolouwu2009}.
However, more and more high-resolution observations
  reveal distinct red asymmetric double-peak profiles
  with weaker blue peaks and stronger red peaks
  \citep[i.e. the so-called red profiles, see e.g.]
  []{mardones1997,thompsonwhite2004,fuller2005,velusamy2008},
  which cannot be understood within the scenario of complete
  collapse models.
The presence of `red profiles' for molecular transitions most
  likely indicates a collapsing central core surrounded by an
  expanding outer envelope
  \citep{loushen2004,thompsonwhite2004,gaolou2009,lougao2011},
  though there have been alternative origins proposed
  (see Gao \& Lou 2010 and Lou \& Gao 2011).

The theoretical framework of self-similar hydrodynamics for
  envelope expansion with core collapse (EECC) was first
  advanced by \citet{loushen2004} for an isothermal gas
  cloud dynamics with spherical symmetry.
This isothermal EECC model was later extended to a more general
  polytropic equation of state (EoS) with the specific entropy
  conservation along streamlines
  \citep[Wang \& Lou 2007;][]{wanglou2008}.

We advance the following scenario for the possible emergence
  of EECC dynamic phase in star-forming molecular clouds.
Both theoretical \citep[e.g.][]{broderick2007,stahleryen2010}
 and observational \citep[e.g.][]{lee2011} works reveal that
 a static core would show expansive or oscillatory motions
 once being perturbed on large scales.
While a molecular cloud is undergoing an expansive or
 oscillatory mode on large scales, its core may begin
 to collapse or contract due to
 nonlinear instabilities under self-gravity.
With the enhancement of such central infalling dynamics
 induced by self-gravity, the collapsed or contracted
 region expands outward in a dynamic manner.
Eventually, the entire cloud starts to collapse once the frontier
 of the collapsed region reaches the outer cloud edge.
Thus, our EECC phase usually exists in the early stage of
 protostar formation, consistent with observations regarding the
 internal dynamic motions of starless cores \citep[][]{lee2011}.

Radiative transfer calculations based on such general polytropic
  EECC shock model have been systematically studied by
  \citet{gaolou2009} and by Lou \& Gao (2011) for the source
  FeSt 1-457 to emphatically demonstrate that the widely
  observed `red profile' signatures in molecular line spectral
  profiles from low-mass star-forming clouds could be sensibly
  viewed as a diagnostic evidence revealing self-similar EECC
  hydrodynamic shock processes therein.
To test the EECC interpretation of `red profiles',
  we investigate low-mass star-forming cloud
  cores with quasi-spherical configurations.
In this paper, we focus on a candidate source of starless cloud
  core L1517B for comprehensive data fitting and comparison.

To achieve high spatial resolutions, star-forming clouds
  with relatively simple internal structures yet without apparent
  envelope disturbances are chosen from the nearby Taurus complex.
After a survey of known cases for star-forming cloud cores with
  red-skewed molecular line profiles in the Taurus complex,
  we choose the cloud core L1517B for a more comprehensive
  investigation as it possibly involves global expansions.
Central spectral line profile observations of molecular
  transitions HCO$^+$($1-0$), HCO$^+$($3-2$), H$_2$CO($2_{12}-1_{11}$),
  H$_2$CO($2_{11}-1_{10}$), CS($2-1$) and CS($3-2$) from this
  candidate source clearly show red asymmetric spectral line
  profiles \citep[e.g.][]{tafalla2004,tafalla2006}, which
  likely reveal the existence of an EECC shock dynamic phase.
Moreover, millimeter and sub-millimeter continuum mappings serve
  as independent constraints on radial density and temperature
  profiles of L1517B \citep[e.g.][]{tafalla2004,kirk2005} for the
  same EECC shock model used for fitting molecular line profiles.

Our motivation is to demonstrate specifically
  that the theoretical explanation of red profiles based on general
  polytropic EECC solutions with collapses, expansions and shocks
  \citep{gaolou2009}, appears grossly consistent with available
  observations for starless cloud core L1517B.
To further constrain and verify the EECC
  shock model, we present the result of a 5 $\times$ 5 (with $50''$
  steps) spatially resolved HCO$^+(1-0)$ molecular line profile
  observation using the 13.7 m telescope of Purple Mountain
  Observatory (PMO) at Delingha in Qinghai Province of China.
We also report the model fitting results of these spatially
  resolved spectra on the basis of the same self-similar EECC
  shock model and show that the EECC shock dynamic phase will
  influence molecular line profiles observed at
  different lines of sight (LOSs) from the cloud,
  rather than simply affects the central spectral
  line profiles alone.
We also predict spatially resolved profiles for
  several other molecular emission lines, sub-millimeter
  continuum at 450$\mu {\rm m}$, and the column density
  distribution of L1517B for future observational tests.

This paper is structured as follows. The background information
  and our motivation are introduced here in Section \ref{sec:level1}.
In Section {\ref{sec:level2}}, we specify
  model parameters and physical properties for L1517B
  using the framework of general polytropic EECC hydrodynamic
  shock model \citep{wanglou2008} without magnetic fields.
Model fittings of central spectra from several
  molecular line transitions as well as spatially resolved
  spectral line profiles for the HCO$^+(1-0)$ transition of
  L1517B are shown and analyzed in Section \ref{sec:level3}.
We also venture to predict other spectral line
  profiles with spatial resolutions.
In Section {\ref{sec:level4}}, we show millimeter and
  sub-millimeter continua as further proof-tests to
  the same EECC shock dynamic model.
Column number density and visual extinctions are predicted
  in Section {\ref{sec:level5}} and we conclude in Section
  {\ref{sec:level6}} with speculations.
Details of analysis are summarized in Appendices A and B for a
  convenient reference.

\section[]{Properties of Starless Cloud
Core L1517B}{\label{sec:level2}}

\subsection{Earlier Studies on Cloud Core L1517B}

The candidate source L1517B as a star-forming cloud core is
  located in the Taurus complex, at an estimated distance
  of $\sim 140$ pc from us \citep[][]{elias1978}.
According to the radio continuum mappings and molecular line
  intensity mappings \citep[][]{tafalla2002,tafalla2006}, cloud
  core L1517B has a radius of $\sim 150''$ (corresponding to
  $\sim 2.1\times 10^4$ AU at a distance of $\sim 140$~pc)
  and a center identified at the right ascension
  $\alpha_{\rm J2000}=04^{\rm h}55^{\rm m}18.8^{\rm s}$
  and the declination $\delta_{\rm J2000}=30^\circ 38'04''$.
Millimeter and sub-millimeter radio continuum data of
  L1517B have been acquired at 1.2~mm wavelength using the
  Institut de Radioastronomie Millim\'etrique (IRAM) 30m
  telescope by \citet{tafalla2002,tafalla2004}, and at
  wavelengths of 850~$\mu$m and 450~$\mu$m using the Sub-millimeter
  Common-User Bolometer Array (SCUBA) on the James Clerk Maxwell
  Telescope (JCMT) by \citet{kirk2005}.
Model analyses of the 1.2~mm continuum have
  been done in two ways in the literature.
One involves a presumed form of the number density radial profile
  $n_0/[1+(r/r_0)^{\alpha}]$ (where $n_0=2.2\times 10^5$ cm$^{-3}$
  is the central number density, $r$ is the radius, $r_0= 35''$ is
  the reference radius and
  $\alpha=2.5$ is the scaling index) plus a constant dust
  temperature $T=9.5$ K \citep[e.g.][]{tafalla2004,tafalla2006}.
The other is a static isothermal Bonnor-Ebert sphere
  \citep{bonnor1956,ebert1955} by \citet{tafalla2002,tafalla2004}.
But the invoked Bonnor-Ebert spheres for clouds appear to
  be dynamically unstable given the inferred physical
  parameters from observations.

These previous analyses offer relevant information for
  L1517B, with a constant
  temperature of $\sim 9.5$~K and a number density of
  $\sim 2.2\times 10^5~{\rm cm^{-3}}$ in the inner region.
Information has also been drawn from the 850 $\mu$m
  continuum observation of L1517B (e.g. Kirk et al. 2005),
  including a temperature estimate of $\sim 10$~K, a full width
  at half maximum (FWHM) diameter of $\sim 0.019-0.028$~pc (by
  a 2D Gaussian fit to the 850 $\mu$m data),
  a central volume number density of
  $\sim 5\times 10^5$~cm$^{-3}$ and a total cloud mass of
  $\sim 1.7~M_{\odot}$ within the 150-arcsec aperture.
Radio continuum mappings on shorter wavelengths
  \citep[850 $\mu$m and 450 $\mu$m in][]{kirk2005} reveal
  more substructures (thus not that spherically symmetric as
  compared to the mapping of 1.2~mm continuum emissions) and may
  be used to constrain theoretical models (see subsection 4.2).
Polarization observations for 850 $\mu$m and 450 $\mu$m emission
  continua are performed using SCUBA by \citet{kirk2006},
  from which a magnetic field strength of $\sim 30~\mu$G
  in this cloud core is inferred.
For the moment, magnetic field is not included
  in our current model formulation.

Various molecular transition lines from
  L1517B have been observed using the IRAM 30 m telescope by
  \citet{tafalla2004,tafalla2006}, and spectral line profile
  fittings based on a static gas sphere close to the center with an
  empirical density profile, a constant temperature and an {\it ad hoc}
  outer ($r>10^4~\rm{AU}$) flow velocity gradient are performed therein.
They also argued for other
  scenarios, such as rotation and pure contraction with
  certain cooling mechanisms near the center to generate
  such asymmetric profiles \citep[e.g.][]{tafalla2004}.
There are inconsistencies in such empirical approach
  from the theoretical perspective as also discussed
  in Gao, Lou \& Wu (2009).
Nevertheless, these model fittings might provide
  gross information for L1517B.
In \citet{tafalla2004}, they declare the existence of internal
  motions of the order of $\sim 0.1~{\rm km~s^{-1}}$ in addition to
  turbulence, which we shall see in the following analysis
  might be related to the cloud systematic infall and expansion motions
  predicted by our general polytropic EECC shock hydrodynamic model.
Another important result is the finding of molecular abundance
  drops towards the cloud center \citep[e.g.][]{tafalla2006}, which
  contributes to molecular line profile structures and will also be
  adopted in spectral line profile fittings of our model analysis
  (see subsection {\ref{sec:level33}} for details).

\subsection{Self-similar EECC Shock Model Parameters}

\begin{table}\label{table:parameter1}

\caption{Six independent physical scaling parameters
 and dimensionless self-similar EECC model parameters}
 \begin{tabular}{@{}lcc}
  \hline
  Scaling  & $k_1^{1/2}$ (km s$^{-0.8}$)
           & $t$ (yrs)\\
  \hline
  Value    & $66.0$
           & $3.0\times10^5$\\
  \hline
 \end{tabular}\\

  \begin{tabular}{@{}lcccc}
 \hline
 Parameters & $\gamma$\ \tablenotemark{a} & $A$
 & $B$\ \tablenotemark{b} & $x_1$\ \tablenotemark{c} \\
 \hline
 Values     & 1.2      & 10.02 &  4.58 & $1.87$\\
 \hline
 \end{tabular}

\tablenotetext{1}{$\gamma$ is the polytropic index
   of the EoS of a general polytropic gas.}
\tablenotetext{2}{Two integration constants $A$ and $B$
  are mass and velocity parameters in the asymptotic solution for
  $x\rightarrow\infty$ [see eqns.
  (\ref{equ:asymptotic solution of v}) in Appendix A]
  adopted as the `boundary condition'.}
\tablenotetext{3}{$x_1$ is the dimensionless upstream
  location of an outgoing shock.}
\end{table}

We adopt the general polytropic self-similar EECC hydrodynamic
  shock model of \citet{wanglou2008} to simultaneously fit both
  the (sub)millimeter radio continuum emissions and several
  available molecular line profiles, especially those profiles
  with red skewed peaks and with spatial resolutions.
The general polytropic model can semi-analytically and numerically
  describe self-similar behaviors of the hydrodynamic evolution,
  i.e., a time-evolving molecular cloud preserves
  its basic structural profiles
  (e.g. the density, temperature, velocity and pressure
  profiles remain similar to their initial profiles),
  of a quasi-spherical polytropic gas under self-gravity
  with the specific entropy conserved along streamlines
  (see Appendix A for basic nonlinear hydrodynamic
  partial differential equations (PDEs),
  the self-similar transformation, notations, and
  definitions therein).{\footnote{To explore the origin
  of ``red profiles'', we actually
  reduce the number of free model parameters in our data
  fittings by requiring a constant specific entropy
  everywhere (i.e. $n+\gamma=2$ is introduced),
  corresponding to the conventional polytropic cloud core.
This constraint is not necessary in the theoretical
  framework of our general polytropic model (Wang \& Lou 2008).
  }
A gas thermal temperature of $\sim 10$ K
  \citep[e.g.][]{tafalla2002,kirk2005} and a typical cloud
  radius of $\sim 0.019-0.028$~pc Kirk et al. (2005)
  are considered for setting pertinent physical
  scalings in our self-similar EECC shock model.
As L1517B appears in an early evolutionary phase
  of protostar formation,
  we may estimate a characteristic infall age of about
  $\sim 3\times 10^5$ yrs for this cloud core according
  to Kirk et al. (2005)
  and \citet{myers2005}.
We also adopt an outer radius of $\sim 2.1\times 10^4$ AU
  for our cloud core model of L1517B, by referring to the observed
  $150''$ radius \citep[e.g.][]{tafalla2002} at an estimated
  distance of $\sim 140$~pc \citep[e.g.][]{elias1978}.

Using these empirical information of
  L1517B, we choose the self-similar model parameters
  and the physical scalings of time $t$ as well as
  sound parameter $k$ as listed in Tables 1 and 2,
  of which the sound parameter $k$ is derived from the
  following empirical length scale
  \begin{equation}
  k^{1/2}t^n\cong 10^4~\textrm{AU}\ .
  \end{equation}
The number density scale of L1517B cloud core for
  hydrogen molecule H$_2$ is taken as [eq
  (\ref{equ:self-similar transformation 2}) in Appendix A]
  \begin{equation}
  \frac{1}{4\pi G\mu m_Ht^2}\cong 10^4\ \textrm{cm}^{-3}\ ,
  \label{equ:density scale}
  \end{equation}
  where we adopt the mean molecular weight $\mu=2.29$
  as used by Harvey et al. (2003).
This number density scale is one order of magnitude smaller
  than the empirical values for the central number density
  $\sim 2-4\times 10^5~{\rm cm^{-3}}$
  \citep[e.g.][Kirk et al. 2005]{tafalla2002},
  as the density radial
  profile of our dynamic model increases rapidly towards the
  core center (see Fig. \ref{fig:physical properties}).
\begin{table}\label{table:parameter2}
\caption{Derived physical scaling $k_2$ and dimensionless
self-similar EECC hydrodynamic shock model parameters}
 \begin{tabular}{@{}lccc}
  \hline
  Scaling & $k_2^{1/2}$ \ \ (km s$^{-0.8}$)\tablenotemark{ a}\\
  \hline
  Value & \ \ $66.8$\\
  \hline
 \end{tabular}

 \begin{tabular}{@{}lccccccccc}
 \hline
 &$n$&$q$\tablenotemark{\ b}&$m_0$\tablenotemark{\ c}&$v_1$&$\alpha_1$
            &$x_2 $&$v_2$&$\alpha_2$\tablenotemark{\ d}\\
 \hline
 &0.8&0&0.039&$-0.17$&$0.70$
       &$1.85$&$0.73$& $1.55$\\
 \hline
 \end{tabular}

 \tablenotetext{1}{The downstream sound parameter
  $k_2=(x_1/x_2)^2k_1$.}
\tablenotetext{2}{Parameter $q\equiv 2(n+\gamma-2)/(3n-2)$ is
  determined by the
  polytropic index $\gamma$ and the scaling index $n=2-\gamma$
  adopted in the self-similar EECC hydrodynamic shock model; this
  corresponds to a constant specific entropy everywhere.}
\tablenotetext{3}{The reduced central point mass $m_0$ is obtained
  by solving the self-similar hydrodynamic equations numerically and
  by matching the central free-fall solution (see Appendix A for details).}
\tablenotetext{4}{Parameters $x_1$ (regarded as independent
  in Table 1), $x_2$, $v_1$, $v_2$, $\alpha_1$ and $\alpha_2$
  are the dimensionless reduced upstream and downstream shock
  locations, velocities and densities respectively,
  and are obtained from hydrodynamic shock jump conditions [see
  eqns.
  (\ref{equ:mass conservation 3})$-$(\ref{equ:energy conservation 3})]
  in the self-similar shock model in Appendix B.}
\end{table}

With these specified scalings, physical variables including
  radius, radial flow velocity, number density and thermal
  temperature of L1517B can be expressed in terms of the
  dimensionless self-similar variable $x$ and reduced
  dependent variables of $x$ using eqns
  (\ref{equ:self-similar transformation 0})$-$(\ref{equ:temperature}),
  viz.
\begin{equation}
r_1=1.06~\times10^4x~\textrm{AU}\ ,\label{equ:radius1}
\end{equation}
\begin{equation}
r_2=1.07~\times10^4x~\textrm{AU}\ ,\label{equ:radius2}
\end{equation}
\begin{equation}
u_1=0.168~v(x)~\textrm{km s}^{-1}\ ,\label{equ:speed1}
\end{equation}
\begin{equation}
u_2=0.170~v(x)~\textrm{km s}^{-1}\ ,\label{equ:speed2}
\end{equation}
\begin{equation}
N_{1}=N_{2}=3.49\times 10^3 \alpha(x)~\textrm{cm}^{-3}\ ,
\label{equ:density}
\end{equation}
\begin{equation}
T_1=7.79~\alpha(x)^{\gamma-1}m(x)^q~\textrm{K}\ ,
\label{equ:temperature11}
\end{equation}
\begin{equation}
T_2=7.98~\alpha(x)^{\gamma-1}m(x)^q~\textrm{K}\ .
\label{equ:temperature12}
\end{equation}
The enclosed mass and the central mass accretion rate [see eqns.
 (\ref{equ:self-similar transformation 2}) and (\ref{equ:accretion
 rate}) in Appendix A] are
  \begin{equation}
M_1=0.336~\frac{m(x)}{(3n-2)}~M_\odot\ ,\label{equ:total mass 11}
\end{equation}
\begin{equation}
M_2=0.345~\frac{m(x)}{(3n-2)}~M_\odot\ ,\label{equ:total mass 12}
\end{equation}
  and
  \begin{equation}
  \dot{M_0}=1.16\times 10^{-6}~m_0~M_\odot \textrm{yr$^{-1}$}\ ,
  \label{equ:accretion rate 1}
  \end{equation}
  where subscripts 1 and 2 refer to physical variables
  on the immediate upstream and downstream sides of
  the outgoing shock front, respectively.
As required by the mass conservation across a shock
  surface, we confirm that $M_1\mid_{x=x_1}=
  M_2\mid_{x=x_2}=2.35M_\odot$ by a direct numerical
  computation with model parameters.
To simplify physical units of variables, we have
  already substituted values of $k,~t$ and $n$
  as listed in Tables 1 and 2
  into eq. (\ref{equ:accretion rate}).

\begin{figure}
\includegraphics[height=0.4\textwidth,width=0.5\textwidth]{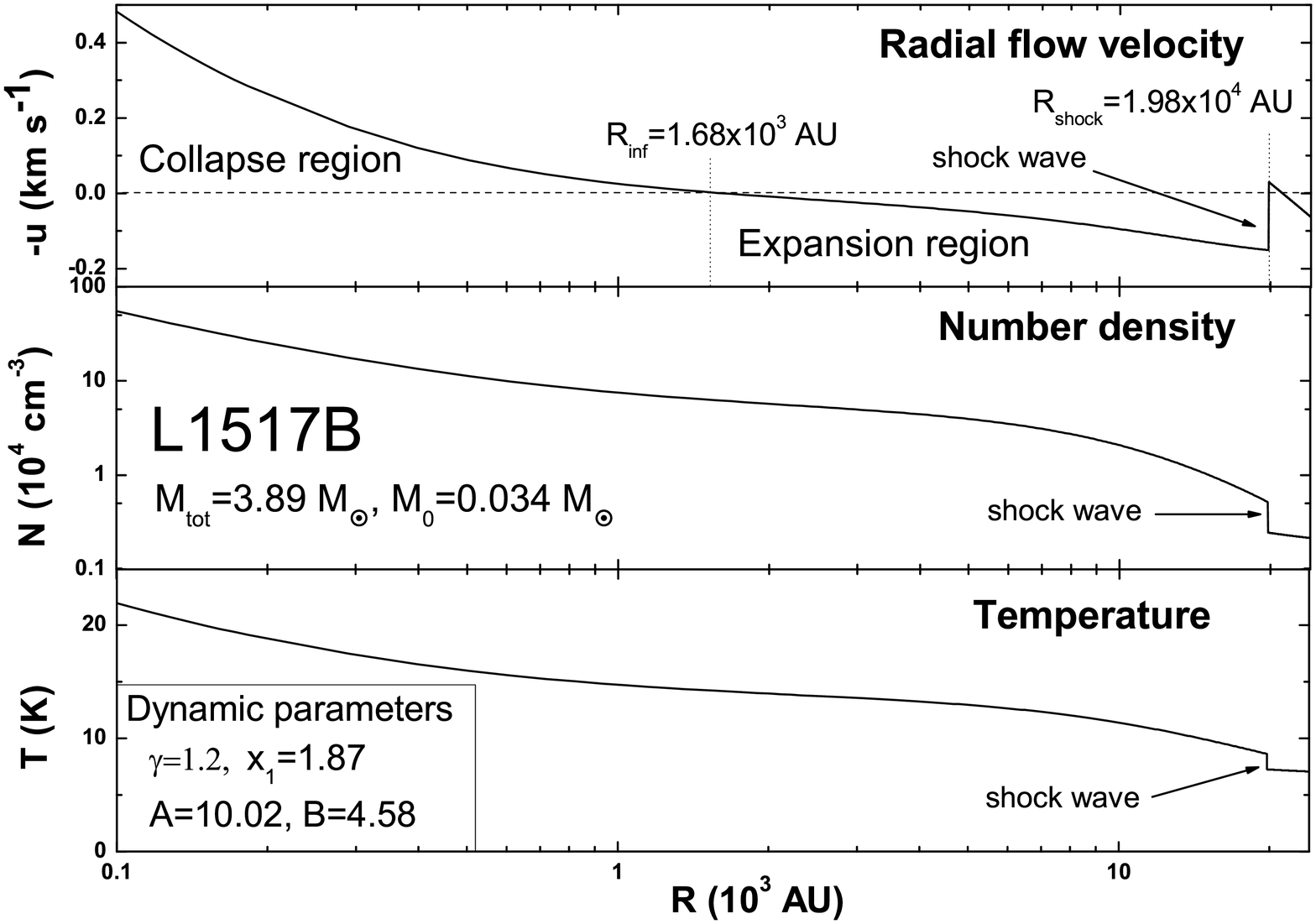}
\caption{\label{fig:physical properties} Structures of
  cloud core L1517B.
From top to bottom are radial profiles of radial flow velocity
  in unit of km s$^{-1}$ (positive values for radial infall),
  number density in unit of $10^4$ cm$^{-3}$ and temperature
  in unit of Kelvin (K), respectively.
The abscissa is radius $R$ in unit of $10^3$ AU
  in a logarithmic scale.
Infall radius $R_{\rm inf}(=1.68\times 10^3~{\rm AU})$ and
 shock radius $R_{\rm sh}(=1.98\times 10^4~{\rm AU})$ expand
 at speeds of $\sim 0.021$ km s$^{-1}$ and $\sim 0.25$ km
 s$^{-1}$, respectively.
Variables exhibit discontinuities
  across the shock front.
Other parameters for this general polytropic self-similar
  EECC shock solution are summarized in Table 3.
The radial density profile of our dynamic polytropic sphere
  with expanding envelope and free-fall collapsing core
  thus consists of two parts connected by an expanding
  stagnation surface, of which the inner part can be
  described as $\rho\propto r^{-3/2}$
  [eq. (\ref{equ:asymptotic solution})] and
  the outer part as $\rho\propto r^{-2/n}$
  [eq. (\ref{equ:asymptotic solution of v})]
  (see the middle panel here).
This broken power law radial density profile is a typical
  feature of starless cores \citep[e.g.][]{Caselli2002}.
  }
\end{figure}

\begin{table*}\label{table:property}
\begin{center}
\caption{Physical properties of star-forming cloud core
  L1517B derived from our polytropic hydrodynamic EECC
  shock model}
 \begin{tabular}{@{}lccccccccccc}
  \hline
  Variables & $M_0$ & $M_{\rm tot}$\tablenotemark{\ a}
            & $\dot{M_0}$ ($M_\odot$ yr$^{-1}$)\tablenotemark{\ b}
            & $R_{\rm inf}$ (AU) & $R_{\rm sh}$ (AU)\tablenotemark{\ c}
            & $u_1$ (km s$^{-1}$)
            & $u_2$ (km s$^{-1}$)\tablenotemark{\ d}\\
  \hline
  Values & $0.034~M_\odot$
         & $3.89~M_\odot$ & $4.54\times 10^{-8}$ & $1.68\times 10^3$ &
           $1.98\times 10^4$
         & $-0.03$
         & $0.12$\\
  \hline
 \end{tabular}

\tablecomments{These physical parameters are grossly consistent
  with those estimated from SCUBA observations by \citep{kirk2005},
  except for the conspicuous feature of the variable temperature profile
  obtained self-consistently from our self-similar EECC hydrodynamic
  shock model, which differs from the static and isothermal assumptions
  in earlier models.
\newline
  (a) $M_0$ and $M_{\rm tot}$ are the central point mass
  and the total cloud mass within a radius $R=2.1\times 10^4$ AU.
\newline
  (b) $\dot{M_0}$ denotes the central mass accretion
rate which decreases with time in our EECC shock model
 (see equation \ref{equ:accretion rate}).
\newline
  (c) $R_{\rm inf}$ is the expanding boundary separating
  the core collapse and envelope expansion regions and
  $R_{\rm sh}$ stands for the shock radius.}
$\!\!\!\!\!\!\!\!\!$ (d) $u_1$ and $u_2$ are the upstream
  and downstream radial velocities across the outgoing
  shock front, with negative values for local inflows
   \end{center}
\end{table*}

All dynamic model parameters in Table 1
  are chosen in the procedure of data fitting of
  molecular spectral lines and dust emission continua
  from observations as described in Sections 3 and 4.
In particular, red profiles, e.g. HCO$^+$($1-0$) and
  HCO$^+(3-2)$ transitions, appear suggestive of an
  expanding envelope in L1517B; this motivates
  us to invoke a self-similar polytropic EECC shock solution.
Meanwhile, in order to fit the molecular spectral line profiles
  of these emissions, we introduce an outgoing shock in the
  EECC model to avoid the outer envelope of this
  molecular cloud expanding too fast.
This shock emerges as an expanding flow rushes into the envelope
  (see the top panel of Fig. \ref{fig:physical properties}).
We have systematically explored a wide range of model parameters
  (including polytropic index $\gamma$, asymptotic behavior
  characterized by mass and velocity parameters $A$ and $B$, and
  location or speed of the outgoing shock) to identify the
  best-fit EECC shock model.
The finally chosen EECC shock model with parameters in Table 1
  does self-consistently and simultaneously fit (sub)millimeter
  emission continua and molecular spectral line profiles
  of L1517B.
Velocity, density and temperature radial profiles at the present
  epoch of the star-forming L1517B described by
  the chosen polytropic EECC shock model are displayed in
  Fig. \ref{fig:physical properties}.
By the very dynamic nature, these radial profiles in Fig.
  \ref{fig:physical properties} qualitatively differ from
  those of both fitting models adopted by
  \citet{tafalla2004,tafalla2006} and Kirk et al. (2005).
In principle, all these model fits should be constrained
  simultaneously by other available observations.

As fitted by the underlying hydrodynamic shock model,
  L1517B appears to involve a
  self-similar core collapse with an envelope expansion
  at a typical outflowing speed of $u_{\rm exp}\sim 0.1$
  km s$^{-1}$ in the radial range of $1.7\times 10^3$ AU
  $\leq R\leq 2.5\times 10^4$ AU, and a concurrent core
  collapse within the infall radius $R_{\rm inf}\cong
  1.68\times 10^3$ AU with a typical infall speed of
  $u_{\rm inf}\sim 0.2$ km s$^{-1}$.
These infall and expansion speeds are in order-of-magnitude
  agreement with the $\sim 0.1$ km s$^{-1}$ internal speed
  estimated in \citet{tafalla2004}.
There is an outgoing shock at $R_{\rm sh}\cong
  1.98\times 10^4$ AU with a travelling
  speed of $\sim 0.25$ km s$^{-1}$.
The central point mass which represents the mass of the
  pre-protostar is $M_0\sim 0.034~M_\odot$ and the total
  cloud mass within a radius of 21000 AU is
  $M_{\rm tot}\sim 3.89~M_\odot$.
Thus, there is not yet a star (in the sense of thermal nuclear
  burning) at the center of L1517B since the inferred mass of
  the pre-protostar is much smaller than the threshold mass
  value ($\sim 0.07~M_\odot$) to initiate nuclear reaction.
Besides, according to the standard equation for the accretion
 luminosity $L_{\rm acc}=GM_\star\dot{M_\star}/R_\star$
 \citep[e.g. equation 2 in][]{kenyonhartman1995}, we can estimate
 an accretion luminosity of about $\sim 0.02L_\odot$ for L1517B,
 which is likely below the current infrared (IR) detection limit.
Moreover, according to \citet{Caselli2002}, starless cores
 are less massive (for example, a statistical mean mass is
 estimated as $\langle M_{\rm tot}\rangle$$\simeq 3M_\odot$,
 while the total mass of L1517B inferred from our model is
 $\sim 3.89~M_\odot$) than cloud cores with nuclear burning
 stars (e.g., $\langle M_{\rm tot}\rangle$$\simeq9 M_\odot$
 in contrast).
Therefore, L1517B can be consistently considered starless,
 rather than a core with a star, as noted before.
Our inferred total cloud mass is twice that derived from
 dust continuum flux densities by Kirk et al. (2005),
 but should still be classified as low-mass
 \citep[$M \leq 8~M_\odot$, cf.][]{mckeeostriker2007}
  star-forming cloud cores.
The ratio of pre-protostar mass to the cloud mass is very low,
  at $M_0/M_{\rm tot}=0.87\%$, which implies that L1517B is
  in an early phase of protostar formation.
The current central mass accretion rate of L1517B obtained
  from eq. (\ref{equ:accretion rate 1}) is
  $\dot{M_0}=4.54\times 10^{-8}~M_\odot$ yr$^{-1}$, giving an
  evolution timescale of $t_{\rm E}\sim M_0/
  \dot{M_0}=7.5\times 10^5$ yr.
From the polytropic model with $n<1$, the mass accretion
  rate decreases with increasing time (see eq.
  \ref{equ:accretion rate}), which appears to be a
  typical feature in low-mass star formation systems
  \citep[e.g.][]{schmejaklessen2004,evans2009}.
Thus the realistic infall age of the cloud should be
  less than the derived evolution timescale $t_{\rm E}$.
Derived physical properties of the pre-protostellar
  core L1517B are summarized in Table 3
  for a convenient reference.

\section[]{Molecular Spectral Line Profiles}{\label{sec:level3}}

Molecular line profile observations reveal rich details for
  physical properties of protostellar cores, including important
  information of the cloud kinematics \citep[e.g.][]{pavlyuchenkov2008}.
We perform spectral line radiative transfer (LRT) calculations
  for molecular line spectra using the publicly available
  numerical code RATRAN \citep[][see website
  http://www.sron.rug.nl/$\sim$vdtak/ratran/ratran.html]{ratran2000},
  which deals with both LRT and non-local thermal equilibrium (non-LTE)
  excitations of molecular energy levels based on Monte Carlo method.
This Monte Carlo code can handle both optically thin
  and thick lines, barring very large optical depths
  (e.g. $\gsim 100$) caused by convergence problems.
We apply the one-dimensional (1D) version of RATRAN
  code for spherical geometry.
We adopt the EECC shock hydrodynamic model with parameters
  specified in Section \ref{sec:level2} for molecular
  spectral line profile calculations.
Here, the radial flow velocity $u$, number density $N$ and
  gas temperature $T$ are all derived self-consistently
  from this EECC model.
The dust temperature $T_{\rm d}$ in the cloud core is assumed to
  be equal to the gas temperature $T$ as expected for high-density
  cloud cores where gas molecules and dusts have sufficiently
  frequent collisional exchanges \citep[e.g.][]{goldsmith1978,goldsmith2001}.
When running the RATRAN code, we divide the spherical cloud
  core into 16 uneven shells with enough accuracy for
  calculations\footnote{Actually, 12 and 14 uneven shells
  have also been tested separately in running the RATRAN
  code and one can sense the gradual convergence to the
  final results with little variance.
  }
  based on the principle that physical quantities
  in each shell do not vary significantly.
We input the number density $N$, gas and dust temperature $T$,
  radial flow velocity $u$ derived self-consistently from our EECC
  shock hydrodynamic model for radiative transfer calculations.
Besides, the temperature of background radiation is the
  cosmic microwave background (CMB) temperature $T_{\rm bg}=2.73$ K.
All molecular line transition data are obtained from the Leiden
  Atomic and Molecular Database \citep[LAMDA, see][]{schoier2005},
  and various molecular abundance ratios with respect to H$_2$
  molecules are presented in Table 4
  and more details of abundance profiles can
  be found in subsection \ref{sec:level33}.

\subsection[]{Red Skewed Double-Peak Line Profiles}{\label{sec:level31}}

\begin{figure}
\includegraphics[height=0.4\textwidth,width=0.5\textwidth]{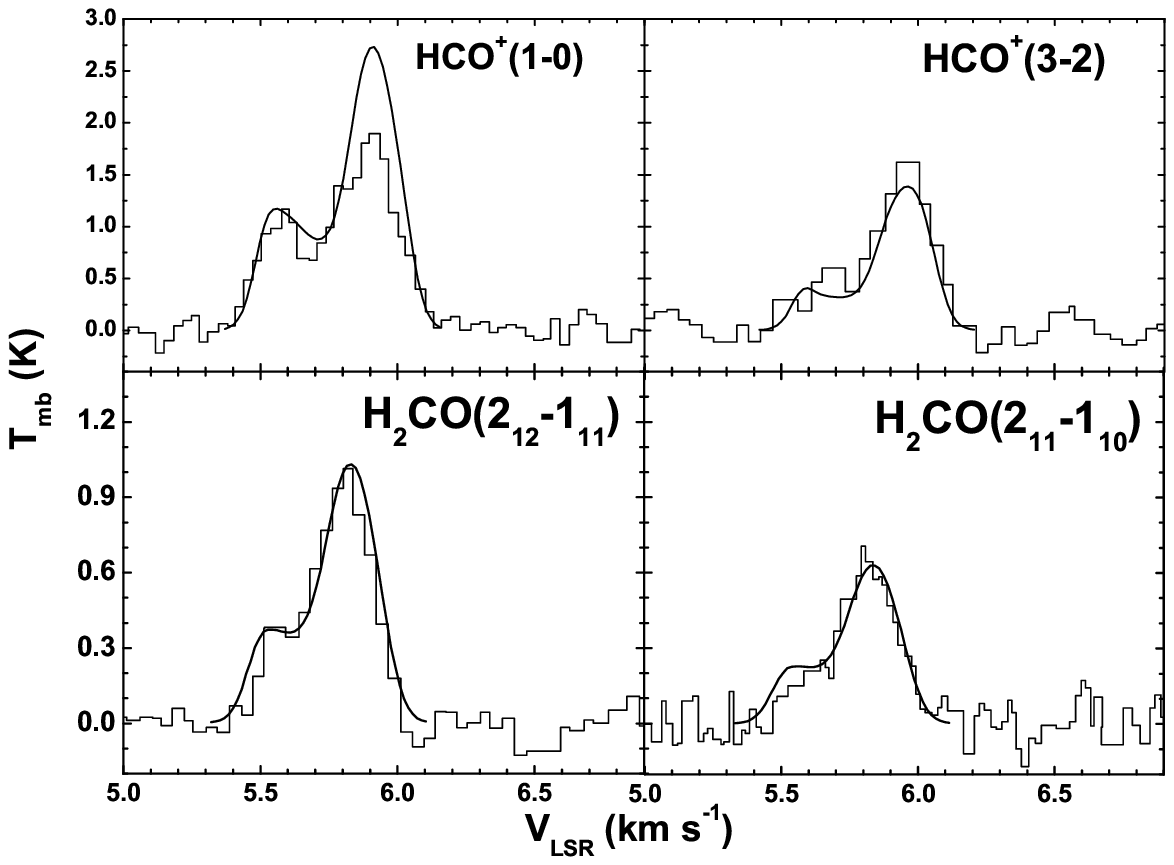}
\includegraphics[height=0.3\textwidth,width=0.5\textwidth]{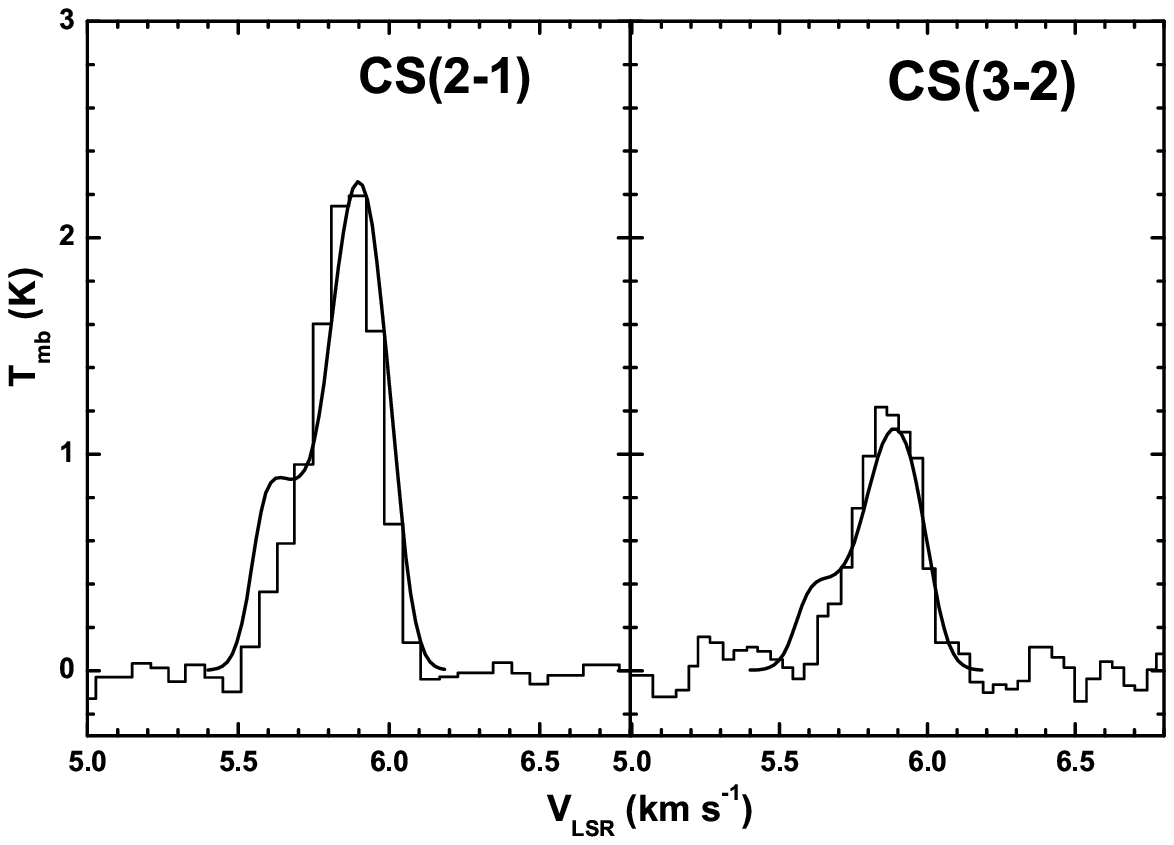}
\caption{\label{fig:red} Central molecular spectral line profiles of
  L1517B with salient red skewed characteristics.
The ordinate stands for the brightness temperature in Kelvin
  and the abscissa represents the LOS velocity component.
Histograms are observational data taken from
  \citet{tafalla2004,tafalla2006} while solid curves
  represent RATRAN LRT computational results
  based on our EECC shock model.
Relevant abundances with respect
  to H$_2$ are listed in Table 4.
Our spectral profile fitting for the molecular transition
  line HCO$^+(1-0)$ shows slightly stronger red peak than observed.
However, spatially resolved observation of this
  line transition reveals that the profiles in some spatial positions
  [see e.g. ($-50'',\ -50''$) in Fig. \ref{fig:all}] exhibit somewhat
  stronger red peaks than our calculation in the central position due
  to non-spherical effects in L1517B.
  For molecular line transition CS($2-1$), the intensity of the
  red peak is more or less the same between the observation
  data and our fitting, while the fitting result of
  \citet{tafalla2004} is lower than the observation data.
These red asymmetric spectral line profiles reveal the presence
  of global expansions for the cloud core envelope.
  }
  \end{figure}

Numerical results (based on our EECC model) of central
  molecular spectral line profiles in comparison with relevant
  observational data from \citet{tafalla2004,tafalla2006} are
  displayed in Figs. \ref{fig:red}, \ref{fig:other molecular lines}
  and \ref{fig:nh3}.
Emission lines of HCO$^+(1-0)$ and H$^{13}$CO$^+(1-0)$ are
  averaged over a beam area of about $50''\times 50''$
  to match with Five College Radio Astronomical Observatory
  (FCRAO) 13.7 m telescope observations in 2001 April
  \citep[e.g.][]{tafalla2006}.
Spectral profiles for the following eight molecular transition
  lines, namely HCO$^+(3-2)$, H$_2$CO($2_{12}-1_{11}$), H$_2$CO($2_{11}-1_{10}$),
  CS($2-1$), CS($3-2$), DCO$^+(3-2)$, SO($23-12$) and SO($34-23$), are
  averaged over a typical beam area of about $20''\times 20''$
  to match with the data acquired using the IRAM 30 m
  telescope in Spain between 1999 October and 2002 November
  \citep{tafalla2002,tafalla2004,tafalla2006}.
Noting that the observational data of the molecular transition
  line CS($2-1$) published in \citet{tafalla2004} actually
  differ from those of \citet{tafalla2002}, we here take the
  observational data of \citet{tafalla2004} in our model
  analysis.\footnote{In our fittings of spectral
  profiles from emission lines of CS, we note that
  the observed intensity of the red peak of CS($2-1$) in
  \citet{tafalla2002} appears too low to simultaneously fit
  the spectral profiles of CS($2-1$) and CS($3-2$)
  transition lines.
We then checked the observations of \citet{tafalla2004}
  and found that Tafalla et al. (2004) have made
  a correction of the relevant data, and the
  new data appears more consistent with our
  model fitting calculations.
The difference between the CS data of Tafalla et al. (2002)
  and Tafalla et al. (2004) may be caused by different
  radio telescopes between FCRAO 14m observations (Tafalla et
  al. 2002) and IRAM 30m observations (Tafalla et al. 2004).}
Following \citet{tafalla2004}, we produce the
  N$_2$H$^+(1-0)$ and N$_2$H$^+(3-2)$ spectra
  by averaging over a beam area of $\sim 26''\times 26''$
  and $\sim 11''\times 11''$, respectively.
As the NH$_3(J,K)=(1,1)$ and $(2,2)$ lines were observed
  simultaneously using the 100 m telescope of the Max Planck
  Institute for Radio Astronomy (MPIRA) at Effelsberg near
  Bonn between 1998 October and 2001 May \citep{tafalla2002},
  we made an average over a $\sim 40''\times 40''$
  beam area to simulate relevant results.
A constant intrinsic line broadening with Doppler b-parameter
  (defined as 1/e the half-width of a line profile in executing
  the RATRAN code)\footnote{Both the thermal ($\Delta v_{\rm T}$)
  and turbulent ($\Delta v_{\rm t}$) components that contribute
  to the line broadening have been included in RATRAN code
  calculations.
$\Delta v_{\rm T}$ is estimated by $\sim (k_{\rm B}T/m)^{1/2}$
  with $T$ consistently obtained from each shell and $m$ is
  the mean particle weight, and
  $\Delta v_{\rm t}=e\times$(Doppler b-parameter).
The total line broadening $\Delta v=[(\Delta v_{\rm T})^2+(\Delta
  v_{\rm t})^2]^{1/2}$.} $\sim 0.08$ km s$^{-1}$ (i.e. with a FWHM
  of $\sim 0.13$ km s$^{-1}$ for a Gaussian line profile) and a
  cloud receding velocity of $u_{\rm cloud}\sim 5.73$ km s$^{-1}$
  are used for all the
  data fitting procedure.

\begin{table}\label{table:molecule}
\caption {
Rest-frame molecular line transition frequencies and molecular
  abundances in reference to H$_2$ molecules}

 \begin{tabular}{@{}lrrcrc}
  \hline
  Molecular Line & Frequency & Abundance 
  \\
  Transitions &  (GHz) & $X_0$ & 
  \\
  \hline
  HCO$^+$(J$=1-0$) & 89.188523 & $1.5\times 10^{-9}$ \tablenotemark{\ (a)} 
                   \\
  HCO$^+$(J$=3-2$) & 267.557619& $1.5\times 10^{-9}$ \tablenotemark{\ (a)} 
                   \\
  H$_2$CO(J$_{K_{-1}K_{+1}}=2_{12}-1_{11}$) & 140.839502 &$5.1\times
  10^{-9}$ \tablenotemark{\ (a)} & 
                   \\
  H$_2$CO(J$_{K_{-1}K_{+1}}=2_{11}-1_{10}$) & 150.498334 &$5.1\times
  10^{-9}$ \tablenotemark{\ (a)} & 
                   \\
  CS(J$=2-1$) & 97.980953 & $2.5\times 10^{-9}$ \tablenotemark{\ (a)} 
                   \\
  CS(J$=3-2$) & 146.969026& $2.5\times 10^{-9}$ \tablenotemark{\ (a)} 
                   \\
  H$^{13}$CO$^+$(J$=1-0$) & 86.754288 &$7.5\times 10^{-11}$ \tablenotemark{\ (a)} 
                   \\
  DCO$^+$(J$=3-2$) &216.112582 & $4.2\times 10^{-10}$ \tablenotemark{\ (a)} 
                   \\
  SO(NJ$=23-12$) &99.299890& $1.4\times 10^{-9}$ \tablenotemark{\ (a)} 
                  \\
  SO(NJ$=34-23$) &138.178670 &$1.4\times 10^{-9}$ \tablenotemark{\ (a)} 
                  \\
  N$_2$H$^+$(JF$_1$F$=101-021$) &93.176258 &$8.0\times 10^{-10}$ \tablenotemark{\ (b)} 
                  \\
  N$_2$H$^+$(JF$_1$F$=344-233$) &279.511811 &$8.0\times 10^{-10}$ \tablenotemark{\ (b)} 
                  \\
  NH$_3$(JKF$_1$F$=1110.5-1111.5$) &23.694501 &$2.5\times 10^{-8}$ \tablenotemark{\ (c)} 
                  \\
  NH$_3$(JKF$_1$F$=2210.5-2211.5$) &23.722680 &$2.5\times 10^{-8}$ \tablenotemark{\ (c)} 
                  \\
  \hline
 \end{tabular}

\tablecomments{
 Frequencies of molecular transitions are obtained
 from the LAMDA database \citep[e.g.][]{schoier2005}.}
 \tablenotetext{1}{
  Molecules with central depletion hole. $X=X_0$ for $r>r_{\rm hole}$,
  and $X=10^{-4}X_0$ for $r<r_{\rm hole}$.
For simplification and for keeping a minimum number of
  parameters in our model construction, we have assumed the
  same radius of abundance hole, $r_{\rm hole}=5.1\times 10^3$
  AU, for all molecular species in the fitting procedure.}
  \tablenotetext{2}{
  Constant abundance without depletion hole.}
   \tablenotetext{3}{ The central abundance enhancement with
Para-$X({\rm NH_3})=X_0({\rm NH_3})[n(r)/n_0]$, where $n(r)$
   is the H$_2$ number density at radius $r$.
We adopt $n_0=2.2\times 10^5\hbox{ cm}^{-3}$ which is
   the same as \citet{tafalla2004} for comparisons.  }
  \end{table}

Optical depths play consequential roles in producing
  asymmetric spectroscopic signatures in molecular line profiles.
Optically thick molecular line transitions of HCO$^+$ and CS
  display deeper self-absorption dips than optically thin
  emissions from molecular line transitions of H$_2$CO,
  which have almost no central dips and show only
  stronger red shoulders (see Fig. \ref{fig:red}).
Moreover, the HCO$^{+}(3-2)$ line transition manifests
 less apparent self-absorption dip in comparison with
 HCO$^{+}(1-0)$ transition.
This contrast is likely due to lower energy level populations
  on J=3 and J=2 in such a cold interstellar medium (ISM)
  environment of $T\sim 10$ K, leading to a lower optical
  depth and source function of the J$=3-2$ line
  transition in HCO$^{+}$ \citep{gaolou2009}.

Emissions from the same molecule but for different energy level
  transitions strongly constrain the underlying polytropic
  hydrodynamic cloud core model, because all the dynamic and
  thermal parameters as well as molecular abundance profiles
  should remain identical in the LRT calculations using the
  RATRAN code.
Fig. \ref{fig:red} shows the spectral profiles of
  HCO$^+$, H$_2$CO and CS each for two distinct
  transitions from central L1517B.
These molecular line spectral profiles present explicit red skewed
  double-peak signatures (i.e. red profiles), consistent with the
  plausible existence of self-similar EECC shock phase
  \citep[Lou \& Shen 2004; Shen \& Lou 2004;]
  []{thompsonwhite2004,gaolou2009}.\footnote{A schematic
  explanation for the origin of red profiles in molecular
  cloud cores can be found in \citet{lougao2011}. }
Such a cloud envelope expansion with core collapse may arise
  from either large-scale cloud radial oscillations
  \citep[e.g.][]{aguti2007} or molecular outflows
  driven by protostar embedded in the core
  \citep[e.g.][]{thompsonwhite2004}.
It has been claimed that the presence of molecular outflows is
  always associated with the evidence of non-Gaussian CO line
  wings \citep[e.g.][]{thompson2004}.
However,
  spectra for
  C$^{18}$O($1-0$), C$^{18}$O($2-1$), C$^{17}$O($1-0$) and
  C$^{17}$O($2-1$) towards L1517B \citep[see figure 8
  of][]{tafalla2004}
  do not manifest such evidence of outflows.
Besides, no spatially separated ``molecular'' outflow
  lobes has yet been detected in this star forming cloud core.
We thus propose that the global envelope expansion with
  central core collapse revealed in L1517B may originate
  from damped acoustic radial pulsations on large spatial
  and temporal scales in molecular cloud cores
  \citep[Lou \& Shen 2004;][]
  {lada2003,keto2006,gaolou2009,lougao2011}.\footnote{
  More explanations for the coexistence of expansion and
  collapse are elaborated in introduction and discussion
  sections.}


\subsection[]{Other Relevant Molecular
Transition Lines}{\label{sec:level32}}

As consistent checks and further constraints,
  we also compute central spectra of isotopologues, namely
  H$^{13}$CO$^+$ and DCO$^+$, of the formyl cation HCO$^+$
  and two distinct line transitions of sulfur monoxide SO
  using the same EECC shock model in
  comparison with observation data of \citet{tafalla2006}.
Fittings of isotopologue emissions with constant abundance ratios
  to each other\footnote{We simply multiply a constant
  ratio to the molecular abundance of HCO$^+$ for
  its isotopologues in our model profile fittings.}
  provide an effective way to investigate the influence
  of abundance patterns on spectral profiles.
As already noted in subsection \ref{sec:level31}, emissions from
  identical molecules with different energy level transitions
  serve as effective evidences to verify our EECC shock model.
In extensive numerical explorations, we adopt molecular
  abundances and isotopic ratios
  which are somewhat different from those used
  in \citet{tafalla2006}, e.g. the SO abundance with respect
  to hydrogen molecule H$_2$ differs from that adopted by
  \citet{tafalla2006} due to the variation in the chosen
  radius of central depletion hole.

There is no conspicuous appearance of red skewed
  double-peak profiles in molecular line
  emissions in Fig. \ref{fig:other molecular lines};
instead, single-peak lines are observed in these
  optically thin lines as expected.
As discussed in subsection \ref{sec:level31}, the absence of
  red asymmetries and self-absorption dips in spectral line
  profiles is attributed to the decrease of molecular level
  populations (which in turn leads to the optical depth
  decrease).
This point is highlighted by comparing the spectral
  line profile of the optically thin H$^{13}$CO$^+(1-0)$
  in Fig. \ref{fig:other molecular lines} with
  that of the optically thick HCO$^+(1-0)$ in
  Fig. \ref{fig:red} [n.b. the optical depth of
  H$^{13}$CO$^+(1-0)$ is $\sim 1/20$ that of HCO$^+(1-0)$}].
Therefore, the combined examination of optically thick and
  thin molecular lines offers an important diagnosis
  to constrain large-scale structures of low-mass star
  forming cloud cores \citep{gaolou2009,lougao2011}.

Compared with observations, our model results in
  Fig. \ref{fig:other molecular lines}, especially those of
  DCO$^+$ and SO, tend to be broader in line profile widths.
Actually, model lines of these molecules
  could be narrowed by reducing their depletion hole sizes\footnote{The
  reduction of the depletion hole (still larger than the collapsing core)
  would bring extra emission contributions from the denser and hotter
  inner region, so we need to decrease the total abundance of these
  molecular species to better fit observations.
Due to the lower abundance, emissions from the outer expanding
  envelope which mainly contribute to the edges of the line spectra
  are weakened, and this leads to narrower spectral line profiles. }.
For illustrations, we have decreased the depletion hole
  radius $r_{\rm hole}$ from $5.1\times 10^3 {\rm AU}$
  to $2.2\times 10^3 {\rm AU}$ for both DCO$^+$ and SO.
Meanwhile, their abundances are reduced to
  $1.5\times 10^{-10}$ and $5\times 10^{-10}$, respectively,
  and then both molecular lines become narrower (see dashed
  curves in Fig. \ref{fig:other molecular lines}).
Physical explanations for the reductions
  of depletion holes are as follows:
The difference between the abundance hole radius of the DCO$^+$ and
  that of the HCO$^+$ results from a central increase in the deuterium
  fractionation caused by the CO depletion, which partly compensates
  the DCO$^+$ freeze out at the inner core \citep[e.g.][]{tafalla2006}.
The non-Carbon-bearing molecule SO does not
  suffer from the depletion of Carbon components, so that its
  depletion hole may differ from those of Carbon-bearing molecules.
Here, SO provides complementary information to emissions from
  Carbon-bearing molecules for the protostar-forming cloud core.

\begin{figure}
\includegraphics[height=0.4\textwidth,width=0.5\textwidth]{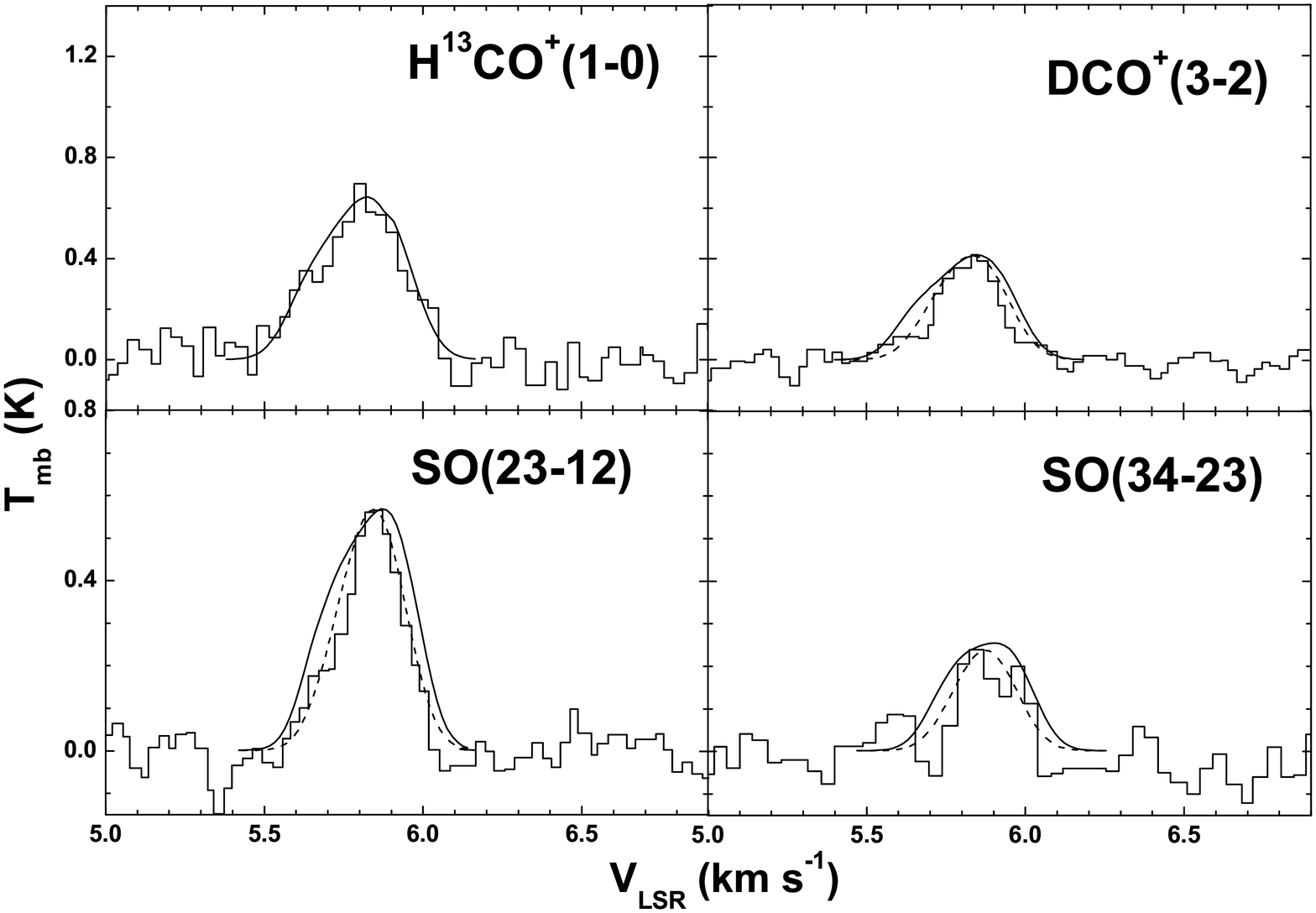}
\caption{\label{fig:other molecular lines} Molecular line profiles
  of central spectra for four transitions
  H$^{13}$CO$^+$($1-0$), DCO$^+$($3-2$), SO($23-12$) and SO($34-23$).
Histograms are observational data taken from \citet{tafalla2006}
  while solid curves represent LRT fitting calculations based on
  the same self-similar EECC shock model.
Solid curves represent the model results with the universal
  depletion hole size
  $r_{\rm hole}=5.1\times 10^3 {\rm AU}$ for all species.
Constant isotopic ratios of
  $^{12}$C/$^{13}$C 
  and H/D 
  (see Table 4 for details)
  are adopted in our model fitting analysis.
Dashed curves displayed in the DCO$^+$ and SO line profiles
  correspond to the simulation results for a smaller depletion
  hole with $r_{\rm hole}=2.2\times 10^3 {\rm AU}$, and lower
  molecular abundances as described in the text.
The absence of red asymmetry and self-absorption dip in these
  molecular spectral line profiles is attributed to the decrease
  of optical depths because of the lower energy level populations
  for these molecular line transitions.
  }
\end{figure}

\subsection[]{Molecules Without Central Depletion Holes}

\begin{figure}
\includegraphics[height=0.4\textwidth,width=0.48\textwidth]{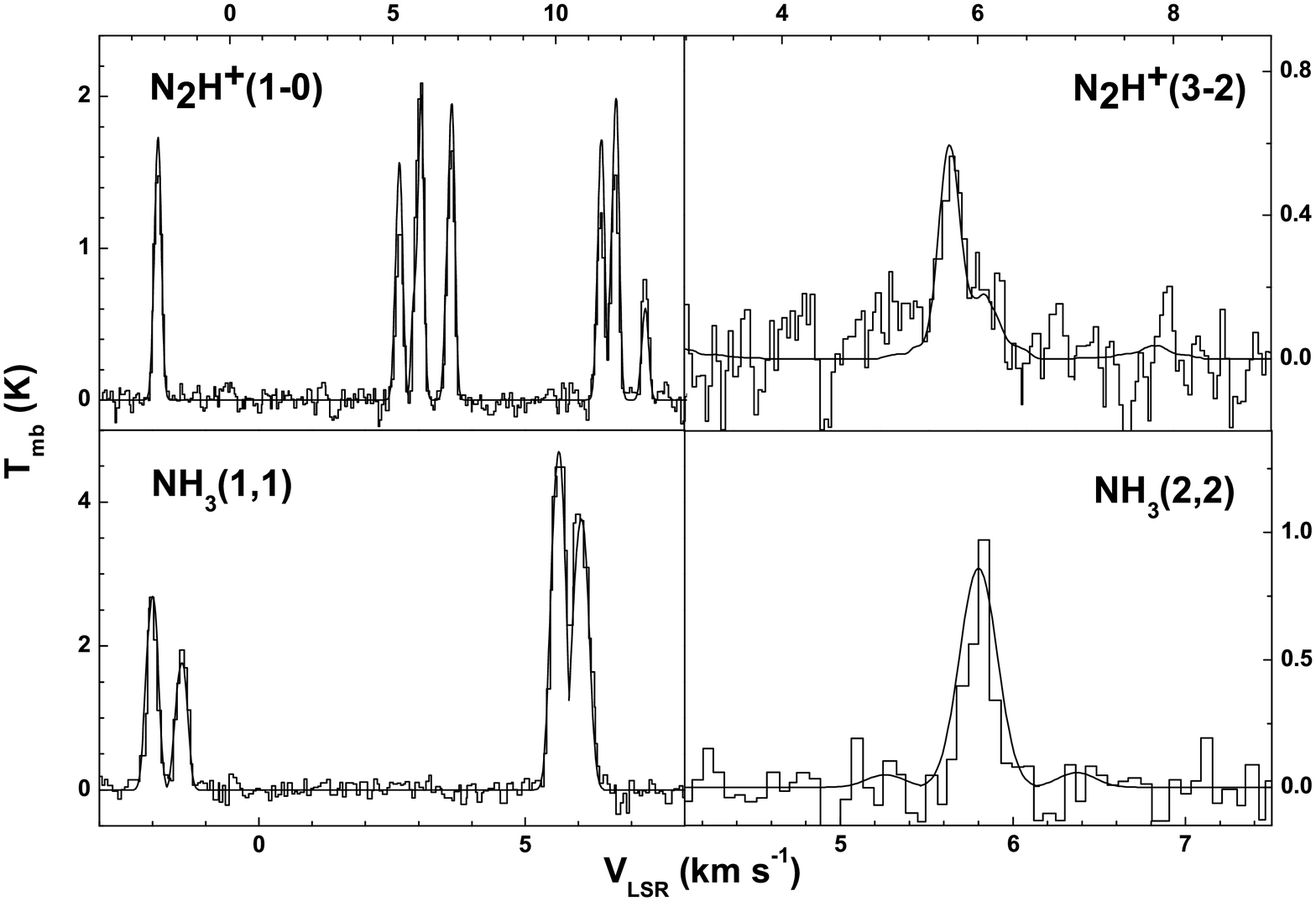}
\caption{\label{fig:nh3} Molecular line profiles from the central
 part of L1517B without depletion holes.
Multi-peaks present in both molecular transitions are caused
 by hyperfine (hf) splittings rather than by the cloud
 core dynamic collapse.
Histograms are observational data taken
 from \citet{tafalla2002,tafalla2004}.
As shown in Table 4, a constant N$_2$H$^+$ (top panels) abundance
 and a central abundance enhancement pattern for NH$_3$ (bottom
 panels) have been adopted in order to fit the observational data.}
\end{figure}

Molecules with central depletion holes offer important diagnostics
  to probe the dynamic structure of the outer molecular envelope
  of L1517B.
Meanwhile, molecular tracers of the central core, such as
  N$_2$H$^+$ and NH$_3$, offer essential information to examine
  the collapsing core, as these molecules do not freeze out onto
  dust grains at typical core densities and they are present in
  the gas phase throughout the core \citep[e.g.][]{tafalla2004}.
As shown in Table 4, a constant N$_2$H$^+$ abundance and a
  variable abundance profile with a central enhancement
  for NH$_3$ have been adopted in our RATRAN calculations.

Unlike other molecular transitions discussed in this
  paper, rotational levels of N$_2$H$^+$ are split
  into multiple hyperfine (hf) components by the
  two nitrogen atoms \citep[e.g.][]{Caselli2002,tafalla2002}.
Such hf splittings make the radiative transfer
  computation much more complicated.
Fortunately, available N$_2$H$^+$ molecular data file
  containing hf structures is provided in LAMDA,
  which simplifies our calculations considerably.
In all, 15 different
  transitions of N$_2$H$^+$ (with 7 distinct frequencies due
  to the overlap of some transitions) for $J=1-0$ and 29
  different transitions (with 26 distinct frequencies)
  for $J=3-2$ are actually involved in their spectra.
We first compute populations of these hf sub-levels and
  related radiative transfer processes with regard to each hf
  transition, and then superpose all the individual hf lines
  to compare with observations (see Fig. \ref{fig:nh3}).
No obvious line broadening deviation from the observed lines has
  been found in our numerical fittings (either in $J=1-0$ or $J=3-2$).
While a significant contracting central region with inward motions
  of the order of $\sim 0.2$ km s$^{-1}$ has been predicted by our
  EECC model, this region is very small as compared to the entire
  collapsing core ($\sim 1/27$ in volume) and is almost negligible
  compared to the entire cloud ($\sim 1/27000$ in volume).
Therefore, this region contributes little to the
  broadening of molecular line profiles.
Such a fairly ``high'' speed of collapsing region is possible
  in a cloud since gas materials immediately above
  the surface of the central dense core (or pre-protostar)
  approach a state of free-fall.
%

The ammonia molecule consists of the so-called ortho and para
  species, which coexist almost independently of each other
  because normal radiative and collisional transitions do
  not change spin orientations \citep[e.g.][]{hotownes1983}.
The observed NH$_3(J,K)=(1,1)$ and $(2,2)$ inversion
  lines arise from para-NH$_3$ \citep[e.g.][]{tafalla2002},
  so we only focus on this particular species in our modelling.
Like N$_2$H$^+$, ammonia NH$_3$ also has hf splittings due to
  the interaction between the electrical quadrupole moment of
  the nitrogen nucleus and the electric field of electrons.
Apart from this major contribution, other weaker interactions,
  including the \textbf{I$_{\rm N}\cdot$ J} (where \textbf{I$_{\rm N}$}
  and \textbf{J} are the nitrogen spin and the total angular momentum
  of ammonia, respectively) and \textbf{I$\cdot$ J} (where \textbf{I}
  is the sum of hydrogen spins) magnetic interactions, as
  well as H-N and H-H spin-spin interactions, further split the
  transition components on the order of
  $\sim 40 {\rm kHz}$ \citep[e.g.][]{hotownes1983}.
Unfortunately, molecular data with hf splittings for radiative
  transfer calculations are not available in LAMDA.
We follow the approach of \citet{tafalla2002} by separating
  the calculation into two steps, namely,
  the solution of the level excitation and the prediction
  of the emergent spectrum.
Using the RATRAN code, we first adopt the ammonia data file
  without hf splittings from LAMDA to obtain the combined
  population of each rotational energy level.
We then compute the population of each hf
  sublevel\footnote{Information about hf sub-levels are
  taken from the NASA-JPL molecular database.
Only the electrical quadrupole hf energy level structure
  has been considered in the data files.
Therefore, 5 distinct components of the transition frequency
  for both $(J,K)=(1,1)$ and $(2,2)$ are taken into account
  in our radiative transfer modelling. }
  by assuming that the sublevels are populated
  according to their statistical weights.
Once populations for the hf sublevels are derived,
  we take the full hf structures into account and
  predict the emergent $(J,K)=(1,1)$ and $(2,2)$ spectra (as shown
  in Fig. \ref{fig:nh3}) by integrating the radiative transfer
  equation along the LOS and summing up each hf
  component\footnote{We compute the Einstein A coefficients
  for different hf line transitions based on equation (9)
  of \citet{pickett1998}.}.
The validity of this approach was
 discussed by \citet{tafalla2002}.
We briefly comment here:
  according to quantum statistics, the population
  ratio between two different levels is given by
  $N_2/N_1=(g_2/g_1)\times e^{-\Delta E/(k_{\rm B}T)}$, where
  $g_{\rm i}$ are the statistical weights (degeneracies) for
  level $i$ ($i$=1,\ 2) and $\Delta E=E_2-E_1$ is the energy
  difference between the two energy levels.
As long as $\Delta E\ll k_{\rm B}T$, i.e. the energy
  difference is far less than the
  energy scale involving the excitation temperature
  (which is usually satisfied when we consider the
  population of hf sub-levels for molecules in starless
  cores), we can then approximately take the population
  ratio as simply $g_2/g_1$.
The excitation temperature and the kinetic temperature
  can be equal under thermodynamic equilibrium, but this
  is not necessarily so in general.

\begin{figure*}
\includegraphics[height=1.05\textwidth,width=0.7\textwidth,angle=270]{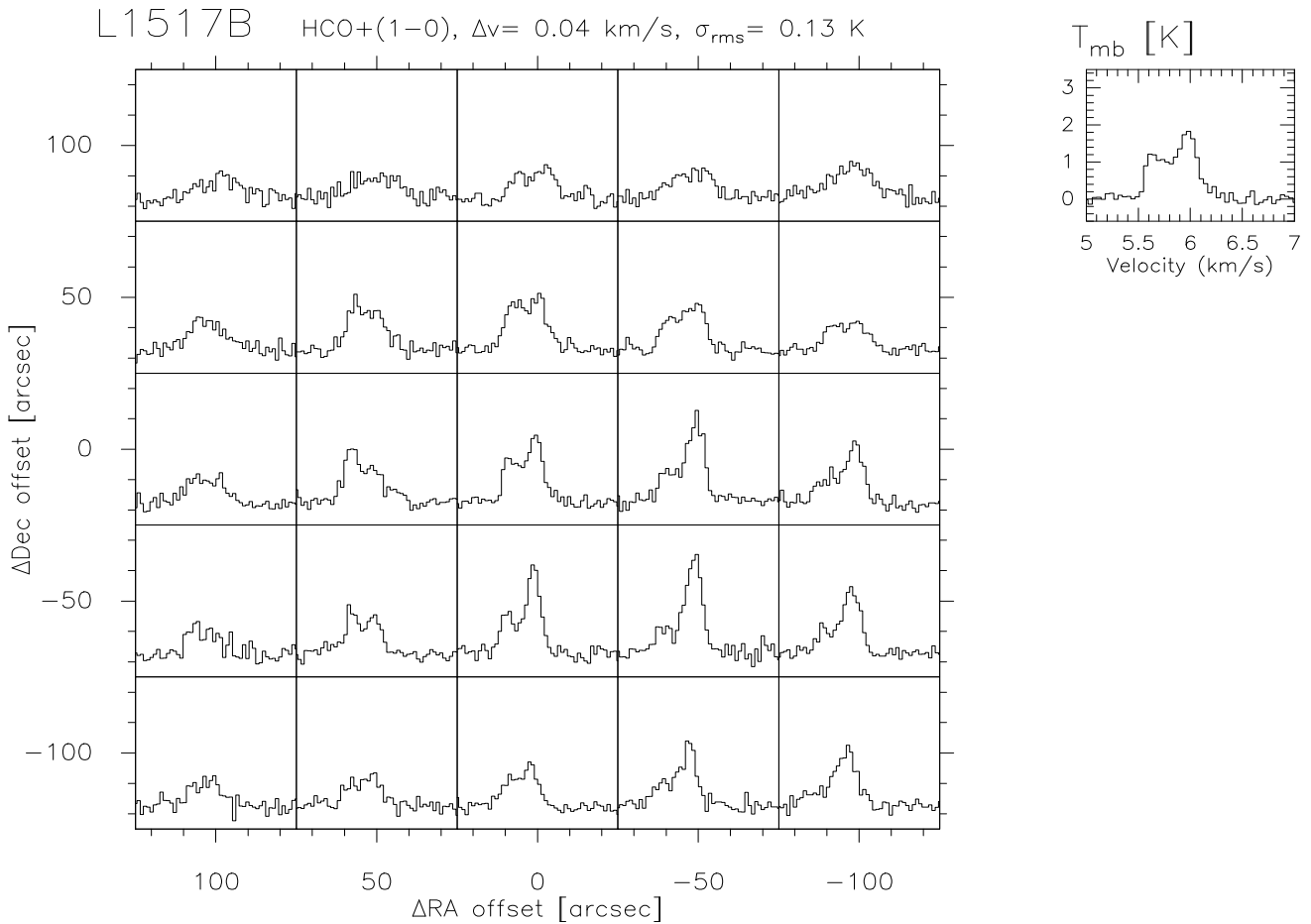}
\caption{\label{fig:all} We show here the grid-map for the
   observed molecular transition HCO$^+(1-0)$ spectra with
   a $50''$ step for L1517B.
The emission intensities are expressed
   in main beam temperature $T_{\rm mb}$
   (using Kelvin as the unit) with elevation corrections
   $f(\theta)=1-0.06(\pm 0.02)\cos\theta$ where $\theta$ is the
   elevation angle of the Delingha telescope, and the efficiency
   of the Delingha 13.7 m telescope is $\eta_{\rm mb}=0.62$.
This $5\times 5$ grip map is centered at $\alpha_{\rm
   J2000}=04^{\rm h}55^{\rm m}18.8^{\rm s}$,
   $\delta_{\rm J2000}=30^\circ 38'04''$ based on the
   central position of the 1.2 mm continuum source of
   L1517B as observed by \citet{tafalla2002}.
The LOS velocity resolution is $\Delta v=0.04$ km s$^{-1}$, and
   the root-mean-square error in the antenna temperature
   $T^*_{\rm A}$ is $\sigma_{\rm rms}=0.083$ K.
The legend panel to the upper right corner
   illustrates the central panel as an example.
The spectral profiles are not perfectly spherically symmetric
   with the eastern spectra slightly shifting towards the blue.
The difference between the eastern and western spectra might be
   due to the rotation of L1517B about its north-south axis.
However, double-peaked red profiles are evident in HCO$^+(1-0)$
   lines across the cloud globule and an average over the same
   distance from the center reveals stronger red peaks in all
   radial positions (see Fig. \ref{fig:spatial resolution}).
In this sense, the presence of red profiles is indicative of cloud
  core expansive motions \citep[see also][]{aguti2007}. }
\end{figure*}

Ammonia NH$_3$ spectral data are usually used to probe
  the central gas kinetic temperature of star-forming
  molecular cores \citep[e.g.][]{tafalla2004}.
The database of dense molecular cores mapped in the $(J,K)=(1,1)$
  and $(2,2)$ transition lines of NH$_3$ was compiled and
  presented by \citet{jijina1999}, who conclude that the
  temperature distribution of molecular cores in Taurus
  complex is very narrow around $\sim 10$ K.
In our polytropic EECC shock model, however, the gas
  temperature gradually rises up towards the center and
  reaches $\sim 15$ K in the collapsing core of L1517B
  as shown in Fig. \ref{fig:physical properties}.
According to Tafalla et al. (2004) and Ho \& Townes (1983),
  the gas temperature of $\sim 10$ K inferred from NH3
  observations relies on the presumed (static) density
  profile, which is an empirical asymptotic power-law
  envelope with a ``flat" central region.
It appears that such a constant $\sim 10$ K temperature reasonably
  fits the gas kinetic temperature for L1517B derived from NH$_3$
  data analysis \citep[e.g. figure 4 of][]{tafalla2004}.
We note, however, that the gas kinetic temperature thus
  determined is inferred directly from the rotational
  temperature $T^{\rm 21}_{\rm R}$ of NH$_3$
  \citep[e.g.][]{walmsley1983,tafalla2004} which relates to
  the relative brightness temperature/population between
  the $(J,K)=(2,2)$ and $(1,1)$ radiative transitions
  \citep[e.g.][]{hotownes1983};
therefore, the gas kinetic temperature thus derived only
  represents an ``averaged" temperature along the LOS,
  not necessarily indicating the actual temperature at
  a certain spatial point.
In other words, it is possible that a radially variable
  temperature profile could also reproduce the NH$_3$
  data as shown by our dynamic model fitting analysis
  (see also Galli et al. 2002 for variable temperature
  profiles in molecular cloud cores).
Noting that the temperature profile of our model reaches
  $\sim 15$ K only in the very inner region ($\sim$ 1000 AU
  from the center\footnote{The gas kinetic temperature for
  L1517B derived from NH$_3$ observations starts from
  $\sim$ 900 AU \citep[see figure 4 of][]{tafalla2004};
  therefore, the ``hot" region in our model occupies
  the inner core and contributes only a very small
  part to the LOS averaged gas kinetic temperature.})
  while remaining around $\sim 10$ K over a wide radial
  range in the outer portion (see Fig.
  \ref{fig:physical properties}), we would expect that
  our radially variable temperature profile may mimic
  the constant LOS temperature distribution of the
  NH$_3$ analysis for L1517B.
Besides, as the temperature distribution of L1517B in
  figure 4 of \citet{tafalla2004} represents LOS averages
  spatially smoothed with 40'', these averages over the
  telescope beam resolution would further flatten the
  temperature distribution and make the whole profile
  closer to a roughly constant $\sim 10$ K distribution.
Moreover, as our EECC model shown in Fig.
  \ref{fig:physical properties} as a whole reasonably
  reproduces both numerous molecular line emissions
  (including those of NH$_3$ and N$_2$H$^+$) and
  (sub)millimeter dust continuum observations of L1517B
  (see Table 4 and Section 4 for details), we conclude
  that this hydrodynamic self-similar EECC model is a
  viable physical description of cloud core L1517B.
Our temperature variation would only
  introduce minor corrections, because the inner ``hot''
  region is fairly small compared with the entire
  molecular cloud ($R_{\rm inf}<0.1R_{\rm out}$).
In other words, a variable temperature profile might
  appear somewhat flattened in this type of inferences
  by LOS NH$_3$ observations.
Noting that our radial temperature profile drops below
  $\sim 10$ K in the outer envelope portion, it is possible
  to produce a 10 K ``averaged'' rotational temperature
  from our EECC model.
Jijina et al. (1999)
  have adopted the
  $T^{21}_{\rm R}$-$T_{\rm K}$ (here $T_{\rm K}$ is the gas
  kinetic temperature) from \citet{walmsley1983} to convert
  the inferred rotational temperature into gas temperature,
  which may cause further uncertainty in the determination
  of cloud temperature since the
  $T^{21}_{\rm R}$-$T_{\rm K}$ relation of \citet{walmsley1983}
  may not be that accurate \citep[e.g.][]{tafalla2004}.
%

\subsection[]{Influence of Molecular Abundance
  Distributions}{\label{sec:level33}}

The spatial distribution of molecular abundances plays a crucial
  role in producing molecular spectral line profiles and thus
  has received considerable attention in the literature
  \citep[e.g.][]{herbst1973,rawlingsyates2001,tsamis2008}.
Also, the depletion of molecular species by adhesion onto
  cold dust grain surfaces has been discovered in central
  regions of star-forming cloud cores
  \citep[e.g.][]{tafalla2002,tafalla2004,walmsley2004}.
We have selected molecular abundance ratios (relative
  to the number of H$_2$ molecules) by referring to
  \citet{tafalla2004,tafalla2006} yet with variations
  to probe molecular distributions in L1517B.
Central molecular depletion holes with very much lower abundance
  ratios [i.e. $\sim 10^{-4}$ of the ratio in the outer cloud
  layers as in \citet[e.g.][]{tafalla2006}] are
  essential in reproducing absolute intensities and relative
  strengths of `blue' and `red' peaks in molecular spectral
  line profiles.
It is possible to treat the radii of abundance holes,
  which can vary for different molecular species in general,
  as extra independent parameters as in \citet{tafalla2006}.
However, in order to focus on exploring the dynamic process
  of L1517B and to reduce the number of independent
  parameters, we have assumed a common size of depletion
  hole for all molecules, except for NH$_3$ and N$_2$H$^+$
  molecular transitions.
Our chosen abundance distributions and the radius of the
  depletion hole $r_{\rm hole}$ used in the data fitting
  are summarized in Table 4
  for reference.
Based on LRT RATRAN calculations, intensities
  of both `blue' and `red' peaks would decrease with the
  increase of $r_{\rm hole}$;
  this is also intuitively sensible.
In fact, the intensity of the stronger `red' peak appears to
  be much more sensitive to the variation of $r_{\rm hole}$
  than that of the weaker `blue' peak does.
This may allow us to estimate the radius of the depletion hole in
  L1517B by carefully calibrating the `blue' to `red' peak intensity
  ratios in pertinent molecular spectral line profiles.


\subsection[]{Spatially Resolved Spectral
Line Profiles}{\label{sec:level34}}

\begin{figure}
\includegraphics[height=0.4\textwidth,width=0.5\textwidth]{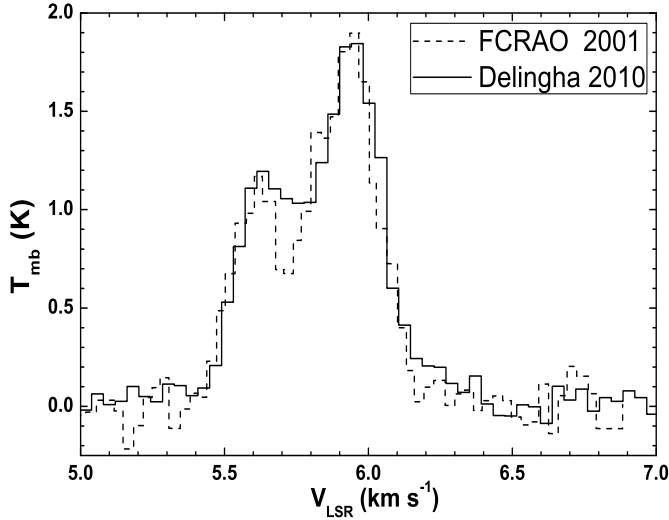}
\caption{\label{fig:compare} A direct comparison between the
  central spectra, which have been generated from the average of
  typically 5 spectra within a $50''$ radius from the core center,
   of the molecular transition HCO$^+(1-0)$ from
  L1517B observed by Delingha 13.7 m millimeter-wave radio telescope
  in mid-April 2010 (solid histogram) and by FCRAO 13.7 m telescope
  in April 2001 (dashed histogram).
The ordinate stands for the main beam brightness temperature
  $T_{\rm mb}$ converted using the beam efficiencies (0.62 for
  Delingha telescope and 0.55 for FCRAO telescope respectively)
  from the respective antenna temperatures $T^{*}_{\rm A}$.
The Delingha observation has a velocity resolution of
  0.04 km s$^{-1}$ as compared with the velocity resolution
  between 0.03 km s$^{-1}$ and
  0.07 km s$^{-1}$ of the FCRAO observation,
  which used the QUARRY array receiver in
  frequency switching mode together with the facility correlator.
With longer integration times for the observation of each
  spatially resolved regions,
  the result from the Delingha observation has relatively
  higher signal to noise ratio (S/N)
  compared with the FCRAO data.
}
\end{figure}

\begin{figure}
\includegraphics[height=0.45\textwidth,width=0.48\textwidth]{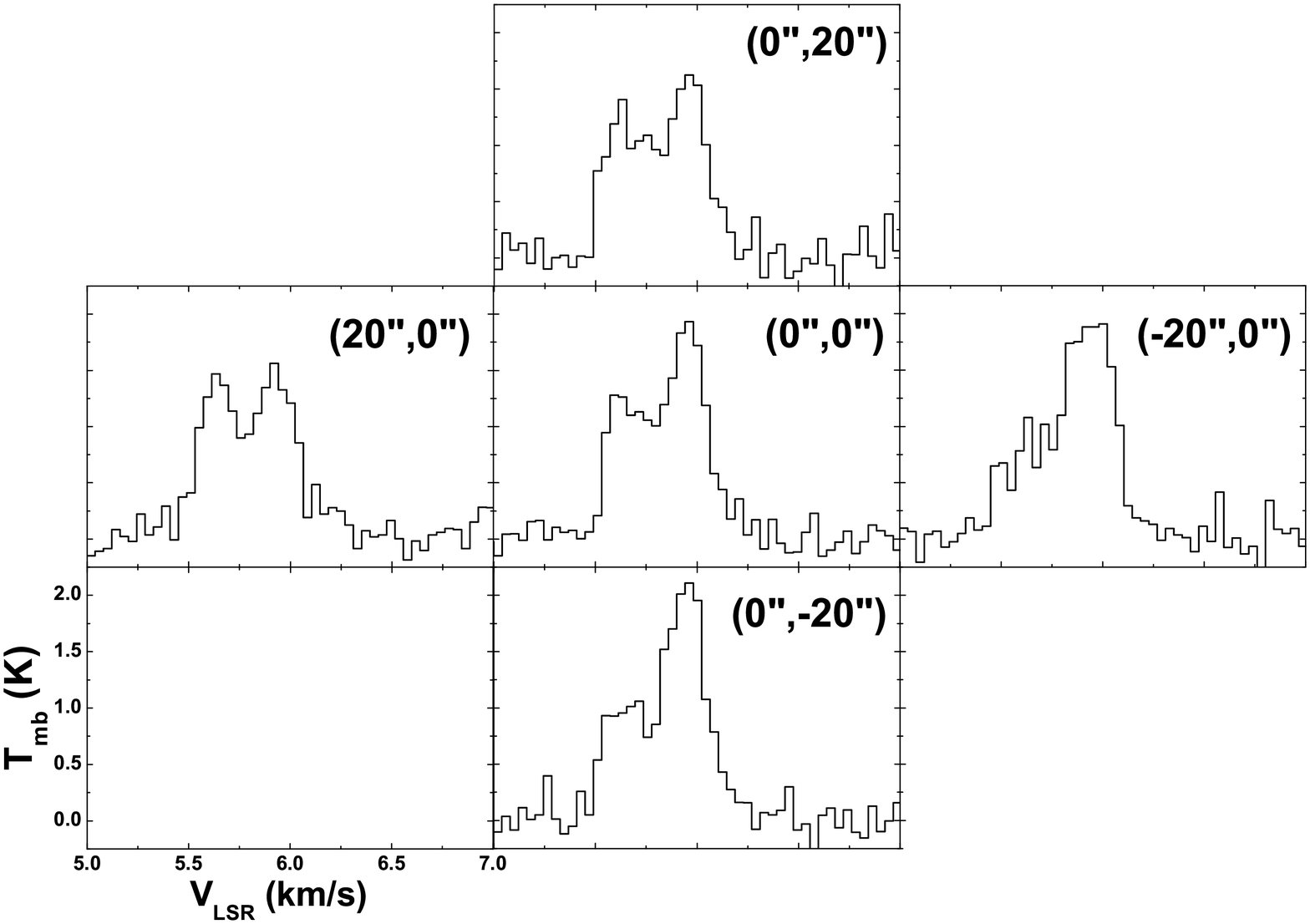}
\caption{\label{fig:20} The HCO$^+(1-0)$ grid-map of four spectral
  profiles at a distance of $\sim 20''$ around the central spectral
  profile plot for L1517B (see Fig. \ref{fig:spatial resolution}
  for the average of these four spectra).
Emission intensities are expressed in the main beam brightness
  temperature $T_{\rm mb}$.
The net integration time for each spectral profile with the
  impact parameter b$=20''$ is 50 minutes, and is 80 minutes
  for the spectral profile at the center.
While these $20''$ spacing spectra overlap previous observations
  with the beam size of $50''$, it is still valuable to examine
  global dynamic properties of L1517B from these spectra,
  as the effect of possible rotation about the north-south axis in
  the inner cloud region appears weaker than in the outer region.
Compared with Fig. \ref{fig:all}, apparent red profiles in nearly
  all these inner spectra (n.b. the red asymmetry is not particularly
  obvious in the eastern spectrum) strongly indicate that this red
  asymmetry is an intrinsic characteristic of L1517B caused by its
  internal motion rather than by rotation.
   }
\end{figure}

Central spectral profile model fittings of various molecular line
  transitions provide an important and effective approach to probe
  both dynamical and thermal structures of pre-protostellar cores.
However, only presenting signatures from the cloud center, central
  spectral profile fittings of molecular lines might still have
  certain ambiguity and uncertainty in determining all pertinent
  physical properties of star-forming cloud cores.
Mapping molecular transition line profiles with high enough
  spatial resolutions, on the other hand, would further constrain
  the large-scale physical conditions within molecular
  clouds \citep[e.g.][]{aguti2007,gaolou2009,lougao2011}.
Here, we report a position-switch observation ($5\times 5$ grids)
  of spectral profiles for the HCO$^+(1-0)$ line transition, which
  is the most prominent diagnostics for dynamic motions within
  L1517B, using the 13.7 m telescope at Delingha in Qinghai
  Province of China and present our model fitting results
  of spatially resolved spectral profiles based
  on the same underlying EECC shock model.
Meanwhile, we predict other pertinent red skewed spectral
  profiles of molecular lines with spatial resolutions
  for future observational tests.

\subsubsection[]{Data Acquisition
and Comparison}\label{sec:level341}

We recently observed L1517B for HCO$^+(1-0)$ at 89.186769 GHz
  using the 13.7 m millimeter-wave radio telescope of PMO at
  Delingha during April 12$-$22, 2010.
The telescope is $\sim 3200$ m above sea level with an
  extremely dry air.
One SIS receiver operating at $\sim 90$ GHz was used in the
  observation together with a FFTS digital spectrometer with
  16384 channels and a working bandwidth of 200 MHz, which
  gives a velocity resolution of 0.04 km s$^{-1}$
  at the frequency of 89 GHz.
The position-switch mode was chosen, with the pointing
  and tracking accuracy within the range of $2''-7''$.
The telescope beam size is approximately 50$''$ at this frequency,
  and we adopted a mapping step size of 50$''$ in observation.
The typical system temperature $T_{\rm sys}$
  is $\sim 250$ K during our observations.
The standard chopper wheel calibration was used during
  the observation runs to get the antenna temperature
  $T^*_{\rm A}$, which has been corrected for the
  atmospheric absorption and telescope elevations.
The beam efficiency of 0.62 is used to convert the
  telescope antenna temperatures $T^*_{\rm A}$ into
  main beam brightness temperatures $T_{\rm mb}$.
The net integration time for most positions is 60 minutes,
  and is 80 minutes for the central position, resulting a
  root-mean-square (rms) noise of $\sigma_{\rm rms}=0.083$ K
  for the measured antenna temperature $T^*_{\rm A}$.

The spectral profile data are processed
  with the standard CLASS software.
Linear baselines are removed from all spectra and an elevation
  correction $f(\theta)=1-0.06(\pm 0.02)\cos{\theta}$
  ($\theta$ is the elevation angle) is adjusted.

We also made an average of five spectra within a 50$''$ radius
  from the cloud core center (i.e. the central spectrum together
  with four spectra that are $50''$ away from the center) to
  directly compare with the result of \citet{tafalla2006}
  obtained by the FCRAO 13.7 m telescope in April 2001.
From the data comparison of Fig. \ref{fig:compare}, we see that
  these two observational results are generally consistent with
  each other with the notable exception of integrated intensities
  around the dip between the two peaks.
This difference may be tolerated within uncertainties of the
  observations and might be
  due to slight discrepancies in telescope
  calibrations or velocity resolutions.
This comparison of data from two independent observations 9
  years apart confirm the validity of both observational
  data from FCRAO 2001 April and Delingha 2010 April.

To examine whether these ``red-profiles''
  are caused by rotation, we have observed four
  spectral profiles at a distance of 20$''$ from the center
  (Fig. \ref{fig:20}) since the rotational effect is much
  weaker in the inner region than in the outer region.
Conspicuous red asymmetric profile characteristics in
  nearly all these spectra exclude the possibility that
  a pure rotation produces these ``red-profiles''.


\subsubsection[]{Model Fittings of Observational Data}

Simultaneous comparisons between model results and
  observational data for molecular line profiles with distinct
  impact parameters $b$ (distance of LOS from the core center)
  is another powerful way to examine the overall dynamical and
  thermal structures of star-forming cloud cores.
As our model is spherically symmetric while
  the spatially resolved observation (Fig. \ref{fig:all})
  exhibits non-spherical effects, we make an average over grids
  of data with the same radial distance from the center.
We carried out observations of four spectra spaced by
  $20''$ from the core center under the same observational
  conditions (yet with 50 minute integration time each) to
  check EECC shock model (Fig. \ref{fig:20}).
All the dynamic parameters as well as the abundance distribution
  and the intrinsic broadening in these simulations are consistent
  with those used in fitting the central spectrum of HCO$^+(1-0)$
  observed by the FCRAO 13.7m telescope (see Section 2).

\begin{figure}
\includegraphics[height=0.5\textwidth,width=0.5\textwidth]{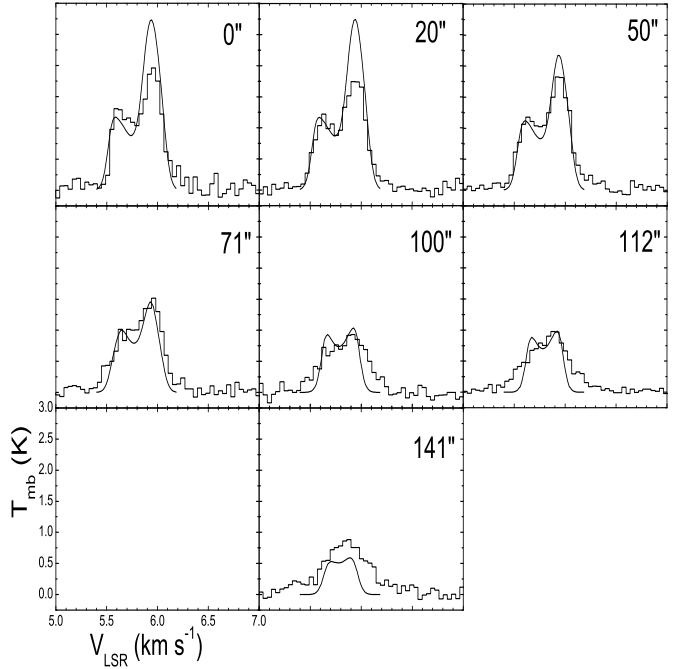}
\caption{\label{fig:spatial resolution}
  Model fittings of spatially resolved spectral line HCO$^+(1-0)$
  observations for L1517B.
Histograms are Delingha observational data produced by the average
  of line spectra with the same distance from the core center.
A telescope efficiency of $\eta_{\rm mb}=0.62$ was used to convert
  the antenna temperature $T^*_{\rm A}$ into the main beam
  brightness temperature $T_{\rm mb}$.
Solid curves present LRT fitting simulations averaged
  over a beam area of $50''\times 50''$ based on the
  same underlying polytropic EECC shock hydrodynamic
  model and molecular abundance pattern.
Spectra correspond to the impact parameter b (distance
  of LOS from the core center) being $0''$, $20''$,
  $50''$, $71''$, $100''$, $112''$ and $141''$, respectively.
The same intrinsic broadening with Doppler b$-$parameter $\sim
  0.08$ km s$^{-1}$ and receding velocity of the cloud core
  $u_{\rm cloud}\sim 5.73$ km s$^{-1}$ as being used previously
  are adopted in our data fitting procedure.
}
\end{figure}

From the model fitting results shown in Fig. \ref{fig:spatial
  resolution}, we see that our computed molecular
  line profiles based on the general polytropic EECC shock hydrodynamic
  model coincide with the observational data.
In comparison with observations, molecular line profiles produced
  from our numerical model calculations show somewhat stronger
  red peaks and weaker blue peaks at the very center
  (Fig. \ref{fig:spatial resolution}).
This likely originates from the spatial asymmetry of the spectral
  line profiles, which might be caused by the possible rotation of
  L1517B about its north-south axis.
The small blue shifts in the eastern spectra (see Fig.
  \ref{fig:all}) will reduce the intensity of the red peaks
  and enhance the blue peaks when we make an average of the
  observational data over the same radial distance.
Meanwhile, these averaged spectra exhibit increasing
  broadening towards the outer envelope while the
  broadening of our numerical results remain invariant.
This phenomenon might be possibly attributed to the incremental
  influence of turbulence with increasing radius, which
  is typical in star-forming clouds \citep[e.g.][]{mckeetan2003}.
In order to reduce the number of free parameters and to
  better understand the dynamics of L1517B, we simply
  assumed that the intrinsic broadenings (which are characterized
  by the Doppler b$-$parameter in the RATRAN code) take the same
  value everywhere in the numerical model fitting calculations.

All these averaged spectral profiles present evident
  red profile characteristics, which are indicative
  of cloud core expansions.
Our fitting results reveal that L1517B is most likely
  undergoing a core collapse and expansion process as
  described by the polytropic EECC shock model.

\subsubsection[]{Several Model Predictions for Cloud Core L1517B}

\begin{figure}
\includegraphics[height=0.30\textwidth,width=0.5\textwidth]{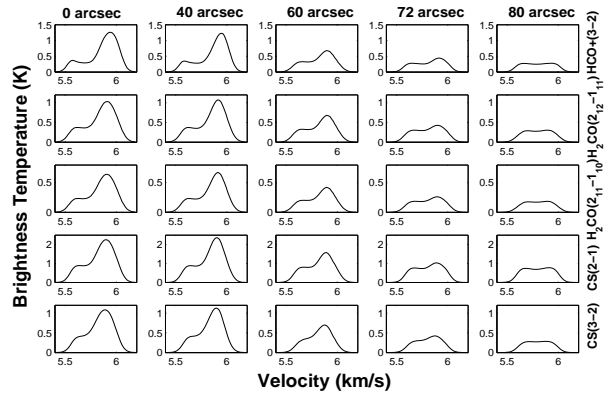}
\caption{\label{fig:space}Red-skewed molecular spectral
  profiles with spatial resolutions for L1517B.
Emissions from HCO$^+(3-2)$, H$_2$CO($2_{12}-1_{11}$),
  H$_2$CO($2_{11}-1_{10}$), CS($2-1$) and CS($3-2$) are
  averaged over a radius of $\sim 20''$
  to simulate the observation of IRAM 30 m telescope.
We present emission spectral profiles with the chosen
  impact parameter $b$ being $0''$, $40''$, $60''$, $72''$ and $80''$
  respectively to reveal the variation of the relative difference
  in strengths between red and blue peaks as well as the overall
  magnitudes.
 }
\end{figure}

As predictions for future observations of
  L1517B, we compute spatially
  resolved relevant molecular line profiles shown in
  Fig. \ref{fig:space} for molecular line transitions
  with red asymmetries in their central spectra.

Our model simulation results in Fig. \ref{fig:space}
  show a gradual disappearance of red asymmetries in spectral
  line profiles and an explicit decrease in overall magnitudes
  as the LOS departs away from the cloud core center.
This variation trend is attributed to the diminution in both cloud
  density and temperature, as well as optical depth \citep{gaolou2009}.
Therefore,
  it would help to assess the influence of optical depths on
  molecular line emissions by making comparisons between model results
  and observations for molecular transitions with spatial resolution.

\section[]{Millimeter and Sub-millimeter Continua of
  Star-Forming Cloud Core L1517B}{\label{sec:level4}}

While model fittings to observed molecular spectral line profiles
  provide an important means to infer physical properties of cloud
  cores, molecular line emissions
  only present under certain
  conditions and some molecules may be frozen onto dust grain
  surfaces in certain regions \citep[e.g.][]{walmsley2004}.
Meanwhile, the radial profiles of flux intensities at
  (sub)millimeter wavelengths from dust grain emissions
  are generally independent of chemical processes, thus can serve as
  an independent observational diagnostics to examine and constrain
  density and temperature distributions in star-forming clouds
  \citep[see e.g.][]{bacmann2000,shirley2000,andre2004}.
Moreover, as the dust density is almost independent of
  depletion holes of molecular abundances, millimeter
  and (sub)millimeter continua would offer an additional
  probe to distinguish different models.
For optically thin dust emissions, the integral form of
  the specific intensity from a spherically symmetric
  cloud globule along a LOS with an impact parameter
  $b$ is given by
  \begin{equation}
  I_\nu(b)=2\int_{b}^{R_{\rm out}}B_\nu[T_{\rm d}(r)]
  \frac{\kappa_\nu(r)\rho(r)\ r}{(r^2-b^2)^{1/2}}dr\ ,
  \label{equ:continuum integration}
  \end{equation}
\citep[e.g.][]{adams1991} where $R_{\rm out}$ is the outer radius
  of a cloud globule, and $T_{\rm d}(r)$, $\rho(r)$, $\kappa_\nu(r)$
  and $B_\nu(T)$ are the dust temperature, cloud mass density,
  specific dust opacity and the Planck function, respectively.

\subsection[]{Radio Continuum Emissions
at 1.2\ mm }{\label{sec:level41}}

The 1.2 mm wavelength radio continuum mapping of L1517B is
  centrally concentrated and appears grossly spherical
  \citep[see figure 1 of][]{tafalla2004}, so the spherical
  symmetry is a reasonable first-order approximation for L1517B.
We use the same EECC shock hydrodynamic model described in
  Section 2 to fit the radial profile of 1.2 mm radio continuum
  emission data \citep[see figure 2 in][]{tafalla2004}.
\citet{tafalla2004} acquired the 1.2 mm radio continuum
  mapping from the IRAM 30 m telescope with a beam size
  of $\sim 11''$.
Therefore, we integrate the intensities from
  model calculations within this beam size.
Parameters of the EECC shock model
  are shown in Tables 1 and 2,
  and our theoretical model fitting to the observed
  radial profile of radio continuum is displayed in
  Fig. \ref{fig:1.2mmcontinuum}.
In the model computations, the dust temperature is assumed
  the same as the gas temperature adopted for
  previous molecular spectral line profile calculations shown
  in the Section 4.
We utilize the dust opacity model proposed by
  \citet{ossenkopf1994} with emissions and absorptions
  included but with scatters ignored.
A typical gas to dust mass ratio of 100 to 1 is
  assumed in our cloud model calculations as usual.
Numerical calculations are carried out using the RATRAN code
  with the spherical cloud divided into
  256 spherical uneven shells.
We have also doubled the number of
  shells (i.e. 512) in our calculations (see the dashed curve in
  the top panel of Fig. \ref{fig:1.2mmcontinuum}) and find that,
  except for a tiny increment at small radii, the result
  generally coincides with that produced by 256 shells.
Therefore, our adoption of 256 shells provides sufficient
  computational accuracy for modeling the continuum emissions.
  Though the density and temperature of the cloud increase
  rapidly near the center, they do not lead to a noticeable
  central peak or a prompt increase in the dust continuum
  since our calculation results have been convolved with
  a telescope beam size of $\sim 11''$ which smooths the
  singularities of density and temperature profiles at
  very small radii.


\subsection[]{Continuum Emissions for 850
$\mu$m and 450 $\mu$m}{\label{sec:level42}}

High-quality sub-millimeter continuum observations
  at wavelengths of 850 $\mu$m and 450 $\mu$m
  reveal more structural complexities
  of pre-protostellar
  cores \citep[e.g.][]{holland1999,shirley2000,dye2008}.
As a consistent check, we compute radial profiles for 850 $\mu$m
  and 450 $\mu$m flux intensities using the same EECC model
  and compare them with the observational data of SCUBA by
  Kirk et al. (2005).
The same opacity model for $\kappa_\nu$ \citep{ossenkopf1994}
  and the gas-to-dust mass ratio of 100 to 1 are adopted in
  these model calculations.

\begin{figure}
\includegraphics[height=0.4\textwidth,width=0.5\textwidth]{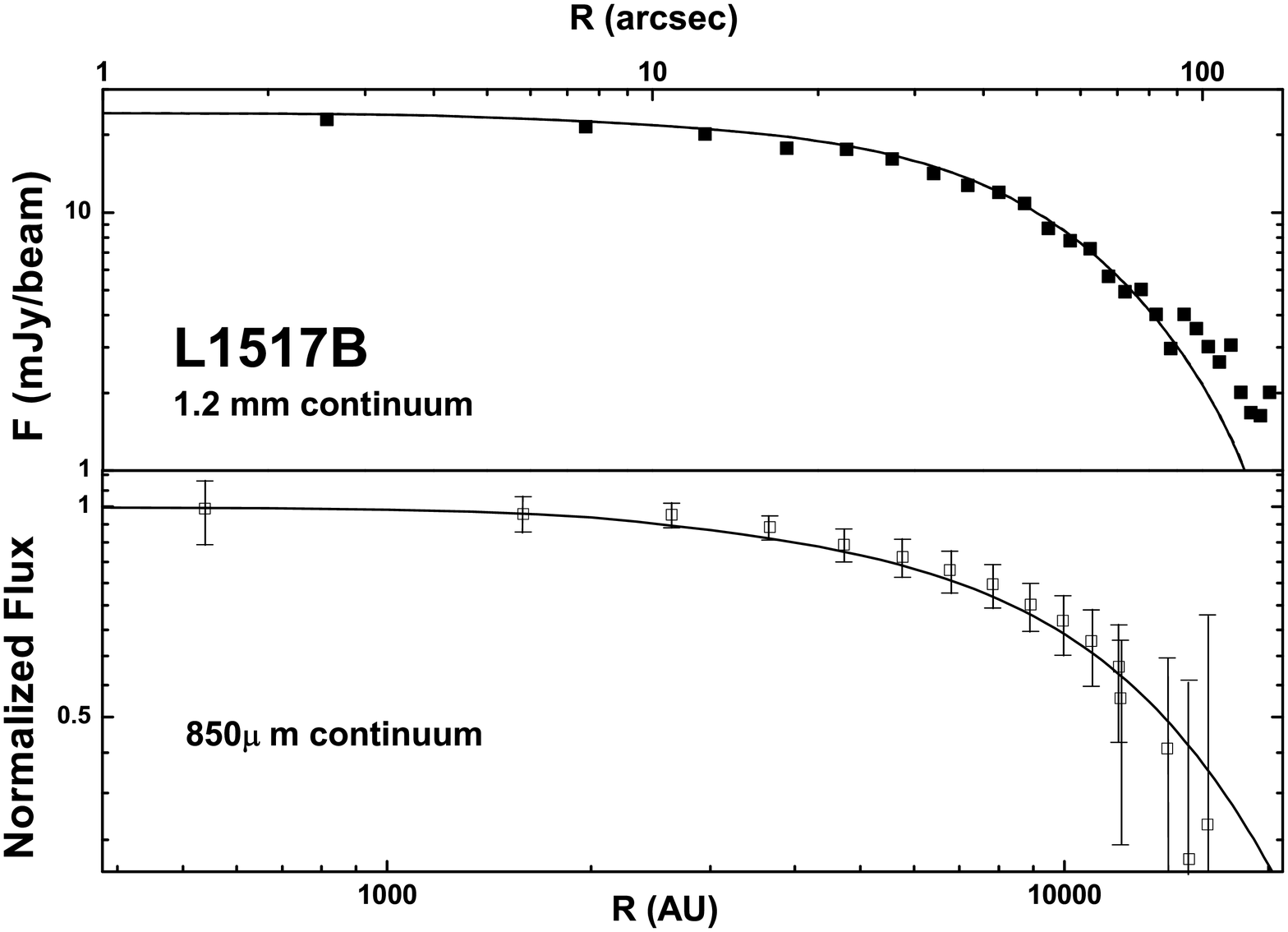}
\caption{\label{fig:1.2mmcontinuum}Radial profiles of the 1.2 mm
 (top) and 850 $\mu$m (bottom) radio continuum emissions from L1517B.
For the 1.2 mm continuum, the intensity scale is in mJy per $11''$
 beam size and the abscissa is the angular distance from the
 cloud core center in arcsec.
Solid squares are data from figure 2
 of \citet[][]{tafalla2004}.
The solid and dashed curves present the simulation
 fitting results based on the density and temperature profiles of
 our EECC shock hydrodynamic model by dividing the cloud into 256
 and 512 shells, respectively.
For 850 $\mu$m, we show the normalized
  radial profile for the flux intensity (per $15''$ beam size) with
  the peak flux density $S_{850}^{\rm peak}\cong$ 155 mJy beam$^{-1}$.
Data points from SCUBA observations (Kirk et al. 2005)
  are shown here in a logarithmic scale at half beam spacings
  with 1$\sigma$ error bars ($\sim 17$ mJy beam$^{-1}$).
The abscissa stands for radius from the center in unit
  of AU on a logarithmic scale.
The scatter of data points, in both the 1.2 mm and 850 $\mu$m
  continua, at the most outer part of the core ($R>80''$, where
  the inconsistency between our simulations and observations
  emerge) is highly likely due to the asymmetry of the core
  \citep[see figure 1 of][]{tafalla2004}.
No conspicuous shock discontinuities are seen for either
  1.2 mm or 850$\rm{\mu m}$ continua,
  since this shock locates at the outer envelope ($\sim
  140~\rm{AU}$)
  where both the temperature and density
  change slightly due to the shock wave.
  }
\end{figure}

In order to be compatible with the JCMT beam FWHM of $\sim 14.8$
  arcsec at wavelengths 850 $\mu$m and $\sim 8.2$ arcsec at 450
  $\mu$m, our numerical calculations are convolved with a
  telescope beam size of $\sim 15''\times 15''$ for 850 $\mu$m
  and of $\sim 8''\times 8''$ for 450 $\mu$m, respectively.
The fitting result of 850 $\mu$m normalized flux density is shown
  in Fig. \ref{fig:1.2mmcontinuum} and the model calculation of
  the 450 $\mu$m normalized flux density radial profile is shown in
  Fig. \ref{fig:450um}.
We find a reasonable agreement of the model results with
  observations of 850 $\mu$m continuum inside $10^4$ AU;
  outside this radius, the quality of data becomes poor
  with considerable error bars.
The 450 $\mu$m flux density profile is grossly consistent in
  intensity with
  the observational data (see figure 3 of Kirk et al. 2005)
  at corresponding radius, with several substructures being ignored.
As we have not included the influence of the central point
  source in our calculations, peak flux intensities of both 850
  $\mu$m ($\sim$ 155 mJy beam$^{-1}$) and 450 $\mu$m ($\sim 790$
  mJy beam$^{-1}$) are somewhat lower than those observed,
  namely $S_{850}^{\rm peak}=170\pm 12$ mJy beam$^{-1}$
  and $S_{450}^{\rm peak}=920\pm 120$ mJy beam$^{-1}$
  (see table 1 of Kirk et al. 2005).
Since the radio observations of 1.2 mm continuum and 850 $\mu$m
  data were carried out by two different telescopes (i.e. the
  IRAM for 1.2 mm and the SCUBA for 850 $\mu$m) and our model
  computational result agrees well with the 1.2 mm continuum of
  dust emissions (see Fig. \ref{fig:1.2mmcontinuum}), the systematic
  difference of our fitting for the 850 $\mu$m continuum profile
  may also be related to differences in calibrations between the
  two telescopes, which could be checked by future direct radio
  observations of 850 $\mu$m continuum using the IRAM 30m telescope.

\subsection[]{Column Number Densities and
Dust Extinctions}
{\label{sec:level5}}

The near-infrared dust extinction provides an
  independent diagnostics to trace the column density in
  molecular clouds without involving the dust temperature
  distributions \citep[][]{alves2001}.
It serves as an important constraint to the density radial
  distribution of our general polytropic EECC shock hydrodynamic
  model, as a complementary yet independent diagnostics to
  various molecular line profiles and dust radio continuum
  emission measurements \citep{lougao2011}\footnote{Because
  the temperature profile is not isothermal (as often assumed
  in the literature) in our EECC model, the column density
  radial distribution cannot be directly mapped out by dust
  (sub)millimeter continuum emissions.
  }.
Moreover, column densities in the low-density outer region of
  molecular cloud, where (sub)millimeter dust continuum emissions
  may not be easy to detect, can still be traced by near-infrared
  dust extinction measurements \citep[][]{kandori2005}.
As an essential and necessary check, we thus compute the column
  density radial profile of H$_2$ molecules by simply integrating
  volume densities from the same EECC shock hydrodynamic model.
Since the column density radial profile is sensitive to the outer
  radius $R_{\rm out}$ of a cloud globule, our model fitting to
  the observed radial profile would identify a proper outer
  radius of the cloud core at $R_{\rm out}=2.5\times 10^4$~AU.
We then invoke the empirical conversion relation
  $N({\rm H}_2)/A_{\rm V}=9.4\times 10^{20}$ cm$^{-2}$mag$^{-1}$
  \citep[e.g.][]{kandori2005,kirk2005}, in which $N({\rm H}_2)$ is
  the H$_2$ molecule column number density and $A_{\rm V}$ is the visual
  extinction, to predict the column density and visual extinction for
  cloud core L1517B using our EECC shock model which
  can reasonably fit other currently available observational data.

\begin{figure}
\includegraphics[height=0.4\textwidth,width=0.5\textwidth]{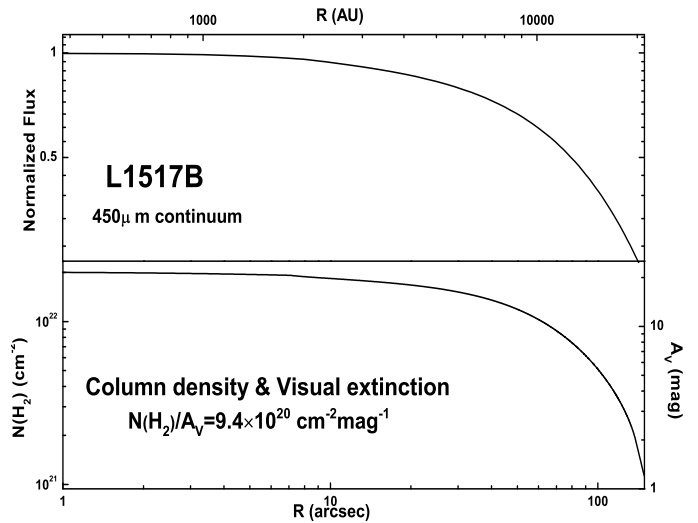}
\caption{\label{fig:450um}Radial profile for the 450 $\mu$m radio
  continuum emissions (top); and column densities for hydrogen
  molecules $N({\rm H}_2)$ and corresponding magnitudes of visual
  extinction $A_{\rm V}$ (bottom) predicted for
  observations of 
  L1517B using the EECC shock model.
For the 450 $\mu$m  continuum, here we show the normalized radial
  profile for the flux density (per $8''$ beam size) with the peak
  flux intensity $S_{450}^{\rm peak}\cong 790$ mJy beam$^{-1}$.
Results are shown in a log-log scale with the abscissa being the
  radial distance in AU.
For the column density and visual extinction profiles, the
  ordinate on the left axis marks the column number density
  distribution $N({\rm H}_2)$ in unit of cm$^{-2}$ and the
  ordinate on the right axis indicates the magnitudes of
  visual extinction $A_{\rm V}$.
The abscissa is the radial angular distance from the cloud core
  center in arcsec.
These results are convolved with a typical observational
  resolution of $\sim 15''$.
}
\end{figure}

In the column density calculation, we convolve numerical
  integration results with a typical beam resolution of $\sim 15''$.
As a comparison and prediction, results of both the H$_2$
  molecule column density and the visual extinction $A_{\rm V}$
  derived from the EECC shock model are
  shown in the lower panel of Fig. \ref{fig:450um}.
For example, the value of central column density predicted by our
  EECC shock model ($\sim 2.2\times 10^{22}$ cm$^{-2}$) is close
  to yet a bit lower than the model result of Kirk et al. (2005)
  [i.e. $N({\rm H_2})_{\rm c}=4\times 10^{22}$ cm$^{-2}$
  with a typical error range of $\pm 20\sim 30$ percent
  (see table 4 of Kirk et al. 2005).
Their central ``column density'', however, instead of being
  directly obtained from infrared dust extinction measurements,
  is actually derived from the 850 $\mu$m radio continuum profile
  modelled by a static isothermal ($T=10$ K ) Bonnor-Ebert sphere
  and a constant dust opacity $\kappa_{850}\equiv 0.01$ cm$^2$ g$^{-1}$.
However, with those estimated parameters, such an isothermal
  Bonnor-Ebert sphere for L1517B turns out to be dynamically unstable.
Besides, their central ``column density'' is model dependent
  without being constrained by other available observations of
  L1517B, so we propose here to test and examine their
  prediction as well as ours for the column density radial profile
  of L1517B by dust extinction observations.

\section[]{Conclusions and Discussion}{\label{sec:level6}}

In this paper, we have systematically investigated the physical and
  chemical (abundance) properties of the starless cloud core L1517B
  from three complementary observational aspects, viz., molecular
  spectral line profiles (some of them are spatially resolved),
  (sub)millimeter continuum emissions and the column density radial
  profile through near-infrared measurements of dust extinction.
We invoke a self-similar polytropic EECC shock model
  with six independent model parameters involved, viz. the
  upstream sound parameter $k_1$, time scale $t$, polytropic
  index $\gamma$, coefficients of asymptotic solution conditions
  $A$ and $B$, and the shock location $x_1$, as summarized in Table 1
  to present reasonable data fittings to various asymmetric `red
  profiles' together with optically thin single peak spectral
  profiles, as well as to 1.2 mm and 850 $\mu$m continuum
  radial profiles from observations
  \citep[Kirk et al. 2005;][]{tafalla2004,tafalla2006}.
In addition, we have recently completed a $5\times 5$ (with a
  50$''$ spacing) spatially resolved spectral profile
  observation of the HCO$^+(1-0)$ line transition using the
  PMO 13.7 m millimeter-wave radio telescope at Delingha
  and made a comprehensive comparison with our model results.
Furthermore, other predicted molecular spectral line profiles
  with higher spatial resolutions are expected to be tested by
  future high spatial resolution spectral observations of L1517B.

The underlying general polytropic self-similar EECC shock
   hydrodynamic model \citep{wanglou2008} provides radial
   velocity, mass density and temperature structures of
   the cloud globule consistently.
Based on the inferred EECC model with a set of fitting
   parameters, we provide physical parameters for L1517B.
(1) The L1517B molecular cloud core is likely evolving in
   an early phase of pre-protostar formation with a small
   central point mass
   of $M_0\sim 0.034~M_\odot$ (incapable of thermal nuclear
   burnings) and a current central mass accretion rate of
   $\dot{M_0}\sim 4.54\times 10^{-8}~M_\odot$ yr$^{-1}$.
(2) The total mass of the cloud core is estimated as
  $M_{\rm tot}\sim 3.89~M_\odot$ revealing that L1517B belongs to
  a low-mass cloud core \citep[e.g.][]{tafalla2002,mckeeostriker2007}.
(3) The cloud core L1517B has an expanding envelope with a typical
  outgoing speed of $u_{\rm exp}\sim 0.1$ km s$^{-1}$
  and a core collapsing at
  $u_{\rm inf}\sim 0.2$ km s$^{-1}$.\footnote{We show
  the velocity profile from our model in Fig.
  \ref{fig:physical properties}; these typical values only imply
  the characteristic magnitudes of expanding and collapsing flow
  velocities, rather than implying the outgoing/infalling region
  has an uniform velocity of this typical value.}
(4) There is also an expanding shock at radius $R_{\rm
  sh}\cong 1.98\times 10^4$ AU with an outgoing radial
  speed of $\sim 0.25$ km s$^{-1}$.
This outgoing shock is initiated when an expanding outflow
  runs into a slowly infalling gas in our dynamic EECC model scenario.
More importantly, we demonstrate that the polytropic EECC shock
  dynamic phase of L1517B with a set of sensible parameters
  can give rise to the manifestation of various observed molecular
  spectral line profiles.

In contrast to the results of \citet{tafalla2004,tafalla2006}
  and Kirk et al. (2005)
  (see section 2.1 for details),
  our general polytropic EECC shock hydrodynamic model simply
  depends on physical properties of L1517B inside the observed
  outer boundary of $\sim 150''$ and self-consistently produces thermal
  as well as dynamic properties of the cloud core in the framework
  of self-similar hydrodynamics to fit the molecular spectral line
  profiles, (sub)millimeter continuum radial profiles and to predict
  the dust extinction property and the column density radial profile
  in the plane of sky.

The uniqueness of our model fitting is generally supported
  by the theoretical work done by \citet{gaolou2009} and
  Lou \& Gao (2011), which discussed the relationship
  between different sets of model parameters and the
  corresponding molecular line spectral profiles.
Instead of other possibilities that might cause red-skewed
  molecular spectral line profiles as noted in the
  introduction of \citet{gaolou2009}, we conclude that the
  `red' asymmetries present in several molecular central
  spectra of L1517B are most plausibly produced by the
  coexistence of core collapse and envelope expansion
  motions in the cloud.
Other scenarios, such as rotation
  and pure contraction without expansion were
  suggested by \citet{tafalla2004}.
However, as discussed in Section \ref{sec:level341},
  spatially-resolved molecular spectra around the core center
  reveal that this asymmetry is most likely an intrinsic
  property of the cloud dynamics, rather than being
  caused by a rotation.
Besides, as the temperature profile in our model
  increases towards the center (a typical characteristics
  for most molecular clouds), it is impossible to produce
  molecular line profiles of red asymmetry from a pure
  contraction \citep[see eq. B4 of][]{gaolou2009}.

As indicated by \citet{broderick2007}, there exist
  radial breathing modes generated from nonlinear evolution
  of acoustic radial pulsations on large spatial and temporal
  scales, with long crossing times in
  protostar-forming molecular cloud cores.
We advanced the following scenario for the
  possible emergence of the EECC phase of L1517B.
The molecular cloud is undergoing an envelope expanding
  phase of the breathing mode when the central contracting
  region collapses due to nonlinear instabilities.
This central infalling dynamics is promoted by self-gravity and
  the collapsed region expands towards the envelope, similar
  to the EWCS
  in \citet{shu1977}.
Eventually, all gas materials would continue to fall towards
  the center to form a proto-stellar core. This scenario is
  consistent with both theoretical \citep[e.g.][]{stahleryen2010}
  and observational \citep[e.g.][]{lee2011} pictures concerning
  the evolution of internal motions in starless cores.

We find in our model analysis that a comparison between optically
  thin and thick molecular spectral line profiles (or between
  different energy level transitions of the same molecule) serves
  as an effective and sensible method to study the influence of
  optical depth variations.
In addition, analyses of the same molecular line transition along
  LOS with distinct impact parameter $b$ from the center
  may further provide an effective approach to examine effects of
  both optical depths and radial variations of cloud physical
  properties (see Section 3.5).
We have prescribed the abundance patterns of different molecules
  with central depletion holes, which has been widely invoked for
  molecular line profiles from star-forming
  clouds as radiative diagnosis
  \citep[e.g.][]{tafalla2002,tafalla2004,walmsley2004}.
Intensities of `red' peaks appear to be much more sensitive
  to the size of a depletion hole, as compared to `blue' peaks.
This may allow us to assess molecular depletions by comparing
  the relative differences between intensities of `red' and
  `blue' peaks in their molecular spectral line profiles.

\acknowledgments

We thank the anonymous referee for suggestions
 to improve the quality of the manuscript.
We would like to thank the hospitality of PMO Delingha
 Observatory and the 13.7~m telescope staff for their
 support during the observation.
D.R. Lu and Y. Sun are acknowledged for assistance in the data
 processing.
We thank J. Yang of PMO for advice and suggestions. This research
 was supported in part
  by the National Undergraduate Innovation Research Program
  091000344 for two consecutive years at Tsinghua University
  from the Ministry of Education.

\appendix

\section{\ \ \ General Polytropic Self-Similar
Hydrodynamic Model}

In spherical polar coordinates $(r,\ \theta,\ \phi)$, general
  polytropic nonlinear hydrodynamic partial differential
  equations (PDEs) for a spherically symmetric molecular
  cloud under the self-gravity and gas pressure force are
  given by
\begin{equation}
\frac{\partial\rho}{\partial
t}+\frac{1}{r^2}\frac{\partial}{\partial r}(r^2\rho u)=0\ ,
\label{equ:equation for density}
\end{equation}
\begin{equation}
\frac{\partial M}{\partial t}+u\frac{\partial M}{\partial r}=0\ ,
\qquad\qquad \frac{\partial M}{\partial r}=4\pi r^2\rho\
,\label{equ:mass conservation}
\end{equation}
\begin{equation}
\rho\bigg(\frac{\partial u}{\partial t}+u\frac{\partial u}{\partial
r}\bigg)=-\frac{\partial p}{\partial r}-\frac{GM\rho}{r^2}\
,\label{equ:momentum conservation}
\end{equation}
\begin{equation}
\bigg(\frac{\partial}{\partial t}+u\frac{\partial}{\partial
r}\bigg)\bigg(\ln\frac{p}{\rho^{\gamma}}\bigg)=0\ ,
\label{equ:entropy conservation}
\end{equation}
where the mass density $\rho$, the radial bulk flow velocity $u$,
  the thermal gas pressure $p$ and the enclosed mass $M$ within
  radius $r$ at time $t$ depend on $r$ and $t$ in general;
  $G=6.67\times 10^{-8}$ dyne cm$^2$ g$^{-2}$ is the
  gravitational constant and $\gamma$ is the polytropic index.
PDEs (\ref{equ:equation for density}) and (\ref{equ:mass
  conservation}) are the two complementary forms of the
  mass conservation.
PDE (\ref{equ:momentum conservation}) is the radial momentum
  conservation of a molecular cloud under the gas pressure
  force and the self-gravity but in the absence of random magnetic
  fields \citep{wanglou2007,wanglou2008}.
PDE (\ref{equ:entropy conservation}) requires the specific
  entropy conservation along streamlines corresponding to a
  general polytropic EoS $p=K(r,\ t)\rho^\gamma$ with $K(r,\ t)$
  formally related to the specific entropy that varies in both
  time $t$ and radius $r$ in general.

To derive an important subset of nonlinear self-similar solutions,
  these hydrodynamic PDEs can be cast into a set of coupled nonlinear
  ordinary differential equations (ODEs) with the following
  transformation \citep{wanglou2008},
\begin{equation}
r=k^{1/2}t^nx\ ,\qquad\qquad {\rm (the\ independent\
self\!-\!similar\ variable\ }x\ {\rm defined)}
\label{equ:self-similar transformation 0}
\end{equation}
\begin{equation}
u=k^{1/2}t^{n-1}v(x)\ ,\qquad\qquad \rho=\frac{\alpha(x)}{4\pi
Gt^2}\ ,\qquad\qquad p=\frac{kt^{2n-4}\beta(x)}{4\pi G}\
,\qquad\qquad M=\frac{k^{3/2}t^{3n-2}m(x)}{(3n-2)G}\
,\qquad\label{equ:self-similar transformation 2}
\end{equation}
where $x$ is the independent self-similar variable combining $r$
  and $t$ in a proper manner, and $v(x)$, $\alpha(x)$, $\beta(x)$
  and $m(x)$ are the dimensionless reduced radial flow speed, mass
  density, thermal pressure and enclosed mass, respectively.
Two constants $k$ and $n$ are the sound parameter and scaling
  index which consistently make $x$, $v(x)$, $\alpha(x)$,
  $\beta(x)$ and $m(x)$ dimensionless and control the spatial and
  temporal scalings of physical variables in a hydrodynamic cloud.
The thermal gas temperature is given by the ideal gas law
\begin{equation}
T\equiv\frac{\mu m_{\rm H}}{k_B}\frac{p}{\rho}=\frac{\mu m_{\rm
H}}{k_B}kt^{(2n-2)}\alpha(x)^{\gamma-1}m(x)^q\ ,
\label{equ:temperature}
\end{equation}
where $k_B$, $m_{\rm H}$ and $\mu$ are Boltzmann's constant, the
  hydrogen atomic mass and the mean molecular weight, respectively.
For a finite $dm(x)/dx$ as $x\rightarrow 0^{+}$ in the central
  free-fall solution \citep[e.g.][]{wanglou2008}, the last expression
  in equation (\ref{equ:self-similar transformation 2}) gives the
  central mass accretion rate as
\begin{equation}
\dot{M_0}=k^{3/2}t^{3(n-1)}m_0/G \label{equ:accretion rate}
\end{equation}
 with $m_0$ being the reduced central enclosed point mass.
For $n$ smaller or larger than 1, this central mass accretion rate
 decreases or increases with increasing $t$, respectively.
For $n=1$, the central mass accretion rate is constant.

Substituting self-similar transformation (\ref{equ:self-similar
transformation 0})$-$(\ref{equ:self-similar transformation 2})
into PDEs (\ref{equ:equation for density})$-$(\ref{equ:entropy
conservation}), we obtain two coupled nonlinear ODEs
\begin{equation}
(nx-v)\alpha'-\alpha v'=-2(x-v)\alpha /x\ , \label{equ:reduced mass
conservation}
\end{equation}
\begin{equation}
\beta'/\alpha-(nx-v)v'=(1-n)v-(nx-v)\alpha/(3n-2)\ ,
\label{equ:reduced momentum conservation}
\end{equation}
with $\beta=\alpha^\gamma m^q$ for $\gamma\neq 4/3$,
 where parameter $q\equiv 2(n+\gamma-2)/(3n-2)$.
These ODEs are related to the mass conservation, the radial
 momentum conservation and the specific entropy conservation
 along streamlines, respectively.
In the limit of $x\rightarrow +\infty$, we have the asymptotic
 self-similar solution to the leading orders as
\begin{equation}
\alpha=A x^{-2/n}+\cdots\ ,\qquad\qquad
v=Bx^{1-1/n}-\bigg[B^2\bigg(1-\frac{1}{n}\bigg)+\frac{nA}{(3n-2)}
+(2n-4)n^{q-1}A^{1-n+3nq/2}\bigg]x^{1-2/n}+\cdots\ ,
\label{equ:asymptotic solution of v}
\end{equation}
(see Lou \& Shi 2011 in preparation for more specific comments on
 the $B^2$ term) where $A$ and $B$ are two integration constants,
 referred to as the mass and velocity parameters, respectively.

In the other limit of $x\rightarrow 0^+$ and to the leading order,
  we have the asymptotic central free-fall solution
\begin{equation}
\alpha(x)=\bigg[\frac{(3n-2)m_0}{2x^3}\bigg]^{1/2}\ ,
\qquad\qquad\qquad v(x)=-\bigg[\frac{2m_0}
{(3n-2)x}\bigg]^{1/2}\ ,
\label{equ:asymptotic solution}
\end{equation}
where the constant $m_0$ is the reduced enclosed point mass,
  representing the dimensionless proto-stellar mass.

By solving coupled nonlinear ODEs (\ref{equ:reduced mass
conservation}) and (\ref{equ:reduced momentum conservation})
  with analytic asymptotic solutions
 (\ref{equ:asymptotic solution of v}) and
 (\ref{equ:asymptotic solution}) as ``boundary conditions", and by
 taking proper care of the sonic critical curve, we can derive
 the radial profiles of velocity, density and thermal temperature
 simultaneously and self-consistently [see \citet{wanglou2008} for
 more details].

\section{Hydrodynamic shock conditions in the self-similar form}

Self-similar shocks may appear in dynamic molecular clouds
  and can be constructed within the framework of a general
  polytropic hydrodynamic model.
For a hydrodynamic shock, we apply the three conservation laws
  in the shock comoving framework, namely conservations of mass,
  radial momentum and energy, across the shock front
\begin{equation}
[\rho(u_{\rm s}-u)]^2_1=0\ , \qquad\qquad [p+\rho(u_{\rm
s}-u)^2]^2_1=0\ ,\qquad\qquad \bigg[\frac{\rho(u_{\rm
s}-u)^3}{2}+\frac{\gamma p(u_{\rm s}-u)}{(\gamma-1)}\bigg]^2_1=0\
,\label{equ:mass conservation 2}
\end{equation}
where $u$, $u_{\rm s}$, $\rho$ and $p$ represent the flow
 velocity, shock front speed, gas mass density and thermal
 gas pressure, respectively.
We use a pair of square brackets outside each expression
 embraced to denote the difference between the upstream (marked
 by subscript `1') and downstream (marked by subscript `2')
 quantities, as has been done conventionally for shock
 analyses \citep[][]{landau1959}.

As the parameter $k$ in the self-similarity
  transformation equation
 (\ref{equ:self-similar transformation 0}) is related to the sound
 speed which are generally different in the upstream and downstream
 sides of a shock, there are two parallel sets of self-similar
 transformation in the two flow regions across a shock.
We set $k_2=\lambda^2 k_1$ with the dimensionless
 ratio $\lambda$ representing this difference.
The relation $x_1=\lambda x_2$ is then required for consistency.
With this similarity scaling relation, hydrodynamic
 shock jump conditions (\ref{equ:mass conservation 2})
 can be readily cast into the following self-similar
 form \citep{wanglou2008}, namely
\begin{equation}
\alpha_1(nx_1-v_1)=\lambda\alpha_2(nx_2-v_2)\ ,\label{equ:mass
conservation 3}
\end{equation}
\begin{equation}
\alpha_1^{2-n+3nq/2}x_1^{2q}(nx_1-v_1)^q+\alpha_1(nx_1-v_1)^2
 =\lambda^2[\alpha_2^{2-n+3nq/2}x_2^{2q}(nx_2-v_2)^q
 +\alpha_2(nx_2-v_2)^2]\ ,\label{equ:momentum conservation 3}
\end{equation}
\begin{equation}
(nx_1-v_1)^2+\frac{2\gamma x_1^{2q}}{(\gamma-1)}
 \alpha_1^{1-n+3nq/2}(nx_1-v_1)^q
 =\lambda^2\bigg[(nx_2-v_2)^2+\frac{2\gamma x_2^{2q}}{(\gamma-1)}
 \alpha_2^{1-n+3nq/2}(nx_2-v_2)^q\bigg]\ .\label{equ:energy
conservation 3}
\end{equation}
These three self-similar shock conditions
 (\ref{equ:mass conservation 3})$-$(\ref{equ:energy conservation 3})
 can be solved explicitly and more relevant details can be found in
 Appendix D of \citet{wanglou2008} in the absence of random magnetic
 fields.





\end{document}